%% file: main.tex
\documentclass[acmsmall]{acmart}
\AtBeginDocument{%
  \providecommand\BibTeX{{%
    \normalfont B\kern-0.5em{\scshape i\kern-0.25em b}\kern-0.8em\TeX}}}

\setcopyright{acmcopyright}
\copyrightyear{2020}
\acmYear{2020}
\acmDOI{}

\acmJournal{JACM}
\acmVolume{37}
\acmNumber{4}
\acmArticle{111}
\acmMonth{8}

\usepackage[usestackEOL]{stackengine}

\usepackage{tikz}
\newcommand*\circled[1]{\tikz[baseline=(char.base)]{
            \node[shape=circle,draw,inner sep=0.7pt] (char) {#1};}}

\usepackage{algorithm}
\usepackage{algorithmicx}
\usepackage{algpseudocode}

\usepackage{amsthm}
\usepackage{multirow}

\theoremstyle{definition}

\begin{document}

\title{PCT-TEE: Trajectory-based Private Contact Tracing System with Trusted Execution Environment}

\author{Fumiyuki Kato}
\email{fumiyuki@db.soc.i.kyoto-u.ac.jp}
\author{Yang Cao}
\email{yang@i.kyoto-u.ac.jp}
\author{Masatoshi Yoshikawa}
\email{yoshikawa@i.kyoto-u.ac.jp}
\affiliation{%
  \institution{Kyoto University}
  \streetaddress{Yoshida-Honmachi}
  \city{Sakyo}
  \state{Kyoto}
  \country{Japan}
  \postcode{606-8501}
}


\begin{abstract}
Existing Bluetooth-based private contact tracing (PCT) systems can privately detect whether people have come into \textit{direct contact} with patients with COVID-19.
However, we find that the existing systems lack {\itshape functionality} and {\itshape flexibility}, which may hurt the success of contact tracing.
Specifically, they cannot detect \textit{indirect contact} (e.g., people may be exposed to COVID-19 by using a contaminated sheet at a restaurant without making direct contact with the infected individual); they also cannot flexibly change the rules of ``risky contact'', such as the duration of exposure or the distance (both spatially and temporally) from a patient with COVID-19 that is considered to result in a risk of exposure, which may vary with the environmental situation.

In this paper, we propose an efficient and secure contact tracing system that enables us to trace both direct contact and indirect contact.
To address the above problems, we need to utilize users' trajectory data for PCT, which we call \textit{trajectory-based PCT}.
We formalize this problem as a \textit{spatiotemporal private set intersection} that satisfies both the security and efficiency requirements.
By analyzing different approaches such as homomorphic encryption, that could be extended to solve this problem, we identify the trusted execution environment (TEE) as a candidate method to achieve our requirements.
The major challenge is how to design algorithms for a spatiotemporal private set intersection under the limited secure memory of the TEE.
To this end, we design a TEE-based system with flexible trajectory data encoding algorithms.
Our experiments on real-world data show that the proposed system can process hundreds of queries on tens of millions of records of trajectory data within a few seconds.

\end{abstract}



\maketitle

\section{Introduction}
\label{sec:1}
Since the beginning of 2020, the emergence of COVID-19 has caused a worldwide pandemic.
Many governments and companies are developing various measures and technologies to prevent the spread of the virus \cite{wang2020response, qin2020dysregulation, salathe2020covid, cho2020contact}.
At present, contact tracing is expected to be a powerful countermeasure for controlling the spread of infection. 
The effectiveness of contact tracing has already been shown by several previous studies \cite{ferretti2020quantifying, Reichert2020PrivacyPreservingCT, tang2020privacy, brack2020decentralized}.
However, conducting effective contact tracing often requires collecting citizens' personal information, such as their locations \cite{privatedata} or telephone numbers \cite{thorneloe2020scoping}, which raises ethical issues and serious privacy violations \cite{jensen2009location}.
Therefore, acceptable private contact tracing (PCT) is urgently needed.

Recently, Bluetooth-based PCT has been intensively studied \cite{troncoso2020decentralized, trieu2020epione, rivest2020pact, becker2019tracking, gvili2020security}. 
Decentralized privacy-preserving proximity tracing (DP3T) \cite{troncoso2020decentralized}, which is an open protocol for PCT that uses Bluetooth low-energy (BLE) beacons, is already being used in applications developed worldwide.
To strongly protect users' privacy, it uses only the contact (proximity) history detected by the BLE beacons.
In DP3T, the applications use the Bluetooth signal of a smartphone to broadcast a random ID that does not include sensitive information such as the user's identity or location, and nearby smartphone devices receive and store the data for a limited time.
Users who are then discovered to be infected with COVID-19 send a report to the server that includes the random IDs they have generated.
Moreover, each user routinely checks to see if the random IDs received from devices they have contacted in the past have been uploaded to the server.
Additionally, there are similar methods for adopting decentralized architecture, such as Epione \cite{trieu2020epione}, the PACT protocol \cite{rivest2020pact}, CEN \cite{becker2019tracking,cengithub} and Google and Apple specifications \cite{gvili2020security}.

However, Bluetooth-based PCT has several limitations in terms of functionality and flexibility.
First, Bluetooth-based PCT detects only \textit{direct contact} (i.e., human-human contact) but cannot detect \textit{indirect contact} (i.e., human-object contact, e.g., when a person visits the same shop shortly after a patient with COVID-19 visited it).
The Centers for Disease Control and Prevention (CDC) in the US showed that it is possible for a person to become infected with COVID-19 by touching a surface or object that has the virus on it and then touching their own mouth, nose, or eyes \cite{cdcreport} --- despite not making direct contact with patients with COVID-19.
Moreover, recent studies \cite{van2020aerosol, xie2020evidence} followed this idea and highlighted the need to trace indirect contact.
Second, Bluetooth-based PCT lacks flexibility in terms of determining the rule of ``risky contact".
Essentially, the rule of risky contact in Bluetooth-based PCT is hard-wired into the Bluetooth device since risky contact is implicitly defined as two devices being in close proximity to each other's signal ranges.
In practice, whether or not risky contact occurs varies with the environmental situation and the nature of the virus.
The rules of risky contact in the case of COVID-19 have been updated as our understanding of the virus \cite{kgwreport} has improved.
For example, in the beginning of the pandemic, professionals believed that transmission took place only through direct human-human contact; however, it was recently argued that airborne transmission should be taken into account \cite{tjtreport, van2020aerosol}.
In addition to the nature of the virus, a recent epidemiological study \cite{pandl2021detection} showed the importance of appropriate selection of the proximity detection range (PDR), which also supports the necessity of flexibility in PCT.
Moreover, recent reviews have pointed out the current PCT application limitations \cite{nanni2020give, 10.1145/3431843.3431845}, which are the inability to detect infections that do not involve direct contact and radio signal limitations in contact detection \cite{ferretti2020quantifying, abbas2020covid}.

In this work, we propose secure and efficient \textit{trajectory-based PCT} to enable both direct and indirect contact tracing.
By comparing the trajectory data between a user and infected patients, we can check whether or not the user visits ``infected locations'' {\itshape within a certain time period}.
The rule of risky contact can be flexibly defined according to the condition of the location and the nature of the virus.
The four requirements for trajectory-based PCT are as follows.

\begin{enumerate}
  \item {\itshape Efficiency}: The central server must be able to handle the query throughput.
  \item {\itshape Security}: A client's trajectory data  must be protected from the server and any other clients. 
  However, nothing about the server-side data is disclosed to the client except the query result. 
  \item {\itshape Flexibility}: The rule of risky contact should be simple to change when necessary.
  \item {\itshape Accuracy}: The server should carefully return accurate results because these results are very sensitive and can significantly affect users.
\end{enumerate}

\begin{figure}[t]
  \includegraphics[width=\linewidth]{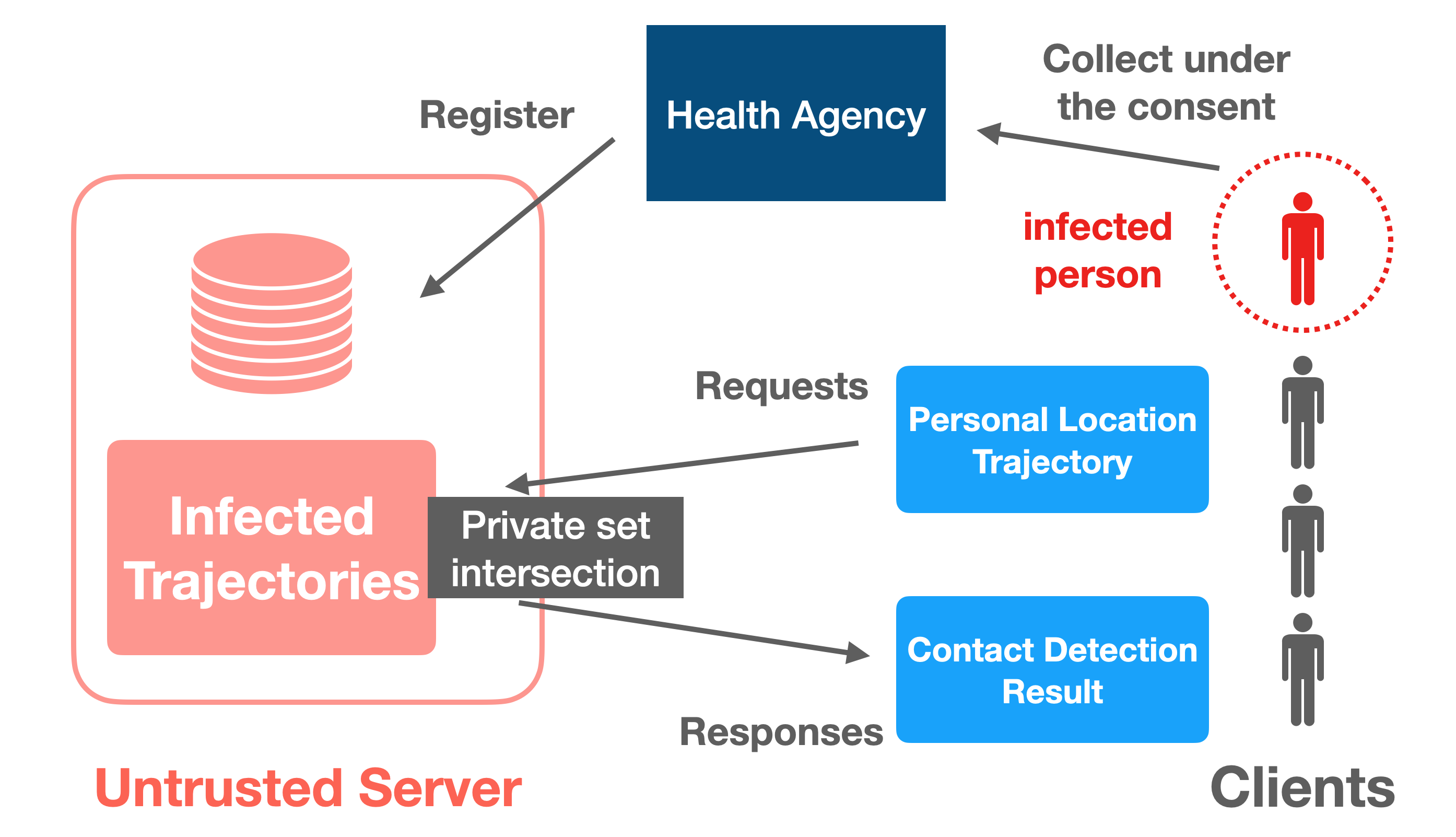}
  \caption{Trajectory-based PCT overview.}
  \label{fig:over_view}
\end{figure}

\begin{table}[t]
\centering
\input{tables/comparison}

\caption{Comparison with existing approaches.}
\label{tab:comparison}
\end{table}

\noindent
As shown in Figure \ref{fig:over_view}, we assume that the health agency (e.g., the government or official healthcare institute) registers the trajectory data of patients confirmed to have COVID-19 (these data are encrypted or released under the consent of the patients) to a server that is untrusted by clients (i.e., queriers).
The server receives queries and encrypted personal trajectories from clients and returns a Boolean value indicating whether there is risky contact or not by computing an intersection between server and client trajectories in a private manner.

Although the problem of trajectory-based PCT is similar to the well-studied problem of a private set intersection (PSI), the existing approaches for the PSI cannot satisfy all four of the abovementioned requirements.
A PSI ensures that two (or more) parties can collaboratively calculate the intersection of their private sets without their private data being disclosed to the other party, only the existing information of the intersection or the result.
However, the existing PSI techniques, based mostly on cryptographic primitives, cannot achieve all of the abovementioned requirements.
The state-of-the-art cryptography-based PSI approaches, such as oblivious transfer \cite{rindal2017malicious} and homomorphic encryption \cite{chen2017fast} have limitations in terms of the {\itshape efficiency}, and there are still performance problems \cite{narayanan2011location, Reichert2020PrivacyPreservingCT} under medium or large workloads, due mainly to the heavy use of time-consuming cryptographic primitives.
Recently, approaches based on secure hardware (such as Intel SGX or ARM TrustZone) have received increasing attention.
Secure hardware enables to make trusted execution environment (TEE) \cite{sabt2015trusted, costan2016intel}, which is used to speed up secure computations on an untrusted party.
Tamrakar et al. \cite{tamrakar2017circle} proposed the first efficient TEE-based PSI. 
It is efficient but does not satisfy our {\itshape accuracy} requirement since it introduces unpredictable error because of the use of probabilistic data structures.
Moreover, {\itshape flexibility} is not considered because their work is based on a general hash function for hash-based dictionary structures.
In our method, we use a flexible encoding hash function to satisfy the requirements.
Thus, we compare their work with ours in Table \ref{tab:comparison}, adding \cite{Reichert2020PrivacyPreservingCT} which is an MPC-based PCT system using trajectories.
Functionality means the capability to detect \textit{indirect contact}.



Our contributions in this paper are threefold.
First, we formulate the problem of trajectory-based PCT. 
We show that our problem is a generalization of the well-studied private proximity testing \cite{narayanan2011location} and PSI.
Our formulation is parameterized for both time and space and can be used in general settings.
We name this formulation the {\itshape spatiotemporal PSI}.
Second, we propose PCT-TEE, a TEE-based efficient algorithm for trajectory-based PCT.
In addition to satisfying the abovementioned requirements, a challenge in designing the TEE-based algorithm is the constraint of secure memory (i.e., enclave) on secure hardware.
We solve these problems by designing a novel trajectory data encoding method, {\itshape TrajectoryHash}, and combining it with {\itshape fast succinct trie} \cite{zhang2018surf}.
{\itshape TrajectoryHash} and the fast succinct trie enable algorithmic flexibility, more efficient compression, and deterministic and fast search performance for a high-speed PSI in a TEE.
Third, we implement the proposed system on Intel SGX and open source the prototype code in GitHub\footnote{https://github.com/ylab-public/PCT}.
Our experiments on real-world datasets show that the proposed system is efficient and effective in practical scenarios.
Specifically, the proposed encoding and data structure compresses the actual trajectory data to one-sixth the size of the Hashmap (also known as a hash table) with the same performance, and as a result, the total execution time is substantially reduced.
Moreover, we show that our system, implemented on a single machine equipped with SGX, can handle hundreds of queries on tens of millions of records of trajectory data in a few seconds.

\noindent
{\bfseries Outline.} 
In Section \ref{sec:2}, as preliminaries, we show that certain features of Intel SGX are related to our system design and discuss the TEE-based PSI and conventional cryptography-based PSI performances.
In Section \ref{sec:3}, we offer the problem statement and formulate the PCT problem.
In Section \ref{sec:4}, we give an overview of our architecture, and in Section \ref{sec:5}, we present the algorithm and trajectory-based data compression.
In Section \ref{sec:6}, we show the experimental results and evaluation.
In Section \ref{sec:7}, we present related works, including related recent PCT applications and our position.
Finally, we provide the conclusions in Section \ref{sec:8}.

\subsection{Relation to the Preliminary Version}
A preliminary version of this article appeared in \cite{kato2020secure}. Compared to the preliminary version \cite{kato2020secure}, there are three significant differences. 
: First, while the preliminary version does not contain clear problem definition for trajectory-based PCT, we formulate the problem as \textit{spatiotemporal PSI}, which can better capture the connection between the geospatial proximity testing problem and PSI problem.
Moreover, we extend the problem to consider the duration of exposure to a virus.
Second, we propose a novel encoding \textit{TrajectoryHash} for compression that preserves trajectory data similarity and compressibility by storing trie-like data structures.
TrajectoryHash is a generalization of our previous encoding and keeps the single trajectory point small.
We analyze closely how the approximation of PCT to PSI by TrajectoryHash causes errors in contact detection.
Third, we evaluate the accuracy of our proposed method.
Our previous work lacks an accuracy evaluation, but the approximation of PCT to PSI causes nonnegligible false negatives.
In this work, we evaluate the accuracy by the extensive empirical experiments, which includes a comparison to the existing work on PCT using trajectory data \cite{Reichert2020PrivacyPreservingCT}.
Based on the results, we discuss how we should apply these PCT techniques to pandemic situations.

\section{Preliminaries}
\label{sec:2}

\subsection{Intel SGX}
Before explaining our system, we introduce the secure hardware used in this paper for ease of understanding our system. 
Below, we focus on Intel SGX, which is a representative implementation of the TEE. 
The architecture and algorithms proposed in this paper can be applied to any other type of secure hardware. 
Intel SGX\cite{costan2016intel} is the extended instruction set of Intel x86 processors, which enables the creation of an isolated TEE, called the {\itshape enclave}.
In addition to powerful server machines, SGX is installed on some PCs.
SGX is also available on some public cloud platforms, such as Azure Confidential Computing, Alibaba Cloud, and IBM Cloud.
We show a brief overview of SGX in the following paragraphs.

The enclave resides in the protected memory region, called the Enclave Page Cache (EPC), in which all programs and data can be unencrypted and quickly processed as well as transparently encrypted outside the CPU package by a memory encryption engine using a secret key that only processor hardware can access.
In other words, SGX adopts a model that considers the CPU package as a trust boundary and everything outside as untrusted.
In this trusted space, accesses from any untrusted software, including the OS/Hypervisor, are prohibited by the CPU, protecting the confidentiality and integrity of the program and data inside the enclave.
Therefore, programs using SGX must use two types of instructions called {\itshape OCALL/ECALL} to invoke functions across trust boundaries under strict control.
These instructions often require too many clock cycles \cite{tian2018switchless}, and does uploading data to the enclave.
This observation is important to improve our system performance.


\noindent
{\bfseries Memory size limitation.}
A challenge in designing algorithms for Intel SGX is the size constraint of the EPC.
The maximum size of the EPC is limited to 128 MB, including 32 MB of metadata for secure management (or 256 MB including 64 MB of metadata in the recent Intel high-end processor\cite{hiendsgx}).
This limitation may be gradually improved but will continue to be a problem regarding hardware and memory-securing performance.
Assume that memory is allocated beyond this memory size constraint. 
In this case, SGX with Linux allows paging with special encryption. 
However, many studies have shown that the performance is greatly degraded by severe overhead \cite{gueron2016memory, gjerdrum2017performance, taassori2018vault}, which is derived from a requirement to preserve confidentiality and integrity even outside the enclave.
Therefore, it is necessary to design an efficient algorithm that works within SGX. 
This is still an important problem, and Kockan et al. \cite{kockan2020sketching} presented a method to overcome the severe memory limitation of TEEs for genomic data analysis.

\noindent
{\bfseries Attestation.} 
SGX supports two types of attestations, i.e., local and remote, which can verify the correct initial state and genuineness of the trusted environment of the enclave from the outside.
In our paper, we focus on remote attestation (RA)\cite{costan2016intel}.
We can request RA to the enclave and receive a report with measurements (e.g., {\itshape MRENCLAVE} and {\itshape MRSIGNER}) based on the hash of the initial enclave state and another environment as a hash chain, which can identify the programs, complete memory layout, and even builder's key information.
This measurement cannot be tampered with.
{\itshape Intel Enhanced Privacy ID} signs this measurement, and {\itshape Intel Attestation Service} can verify the correctness of the signature as a role of a trusted third party.
In addition to verifying the SGX environment, secure key exchange between the enclave and remote client is performed within this RA protocol.
Therefore, after applying the protocol, we can communicate over a secure channel with a remote enclave by a fast encryption scheme such as AES-GCM, and finally, we can safely perform a confidential calculation in the remote enclave.
Our system utilizes this primitive for private computation.

\subsection{Private Set Intersection}
\label{sec:2-2}
The PSI is a well-studied and important problem.
The PSI refers to a setting where multiple parties each hold a set of private sets and wish to learn the intersection of their sets without revealing any information except for the intersection itself.
The existing main approach is to use cryptographic primitives, which are summarized as follows.
We can classify the conventional approaches into two categories; methodology and security model.

Regarding the former, first, there are methods based on the commutative properties of the Diffie–Hellman (DH) key exchange \cite{de2010linear}.
They require computing the polynomial interpolation, which requires a high computational cost.
Huang et al. \cite{huang2012private} described a {\itshape garbled circuit}-based approach.
Their proposed SCS circuit family improved the efficiency at that time.
This approach is similar to the secure-hardware-based approach described later in terms of leveraging a general-purpose secure computation.
{\itshape Oblivious transfer} (OT) \cite{otbased2014benny, rindal2017malicious, pinkas2018scalable} is one of the most promising approaches.
While it is generally used for semi-honest adversaries, \cite{rindal2017malicious} extended the OT method to a malicious adversary using the {\itshape dual execution technique} \cite{mohassel2006efficiency}.
{\itshape Homomorphic encryption} (HE) \cite{chen2017fast, chen2018labeled} is suitable for an unbalanced setting where the server-side data are large and the client-side data are small, because it can replace the oblivious pseudorandom function in the OT-based approach with leveled fully HE and substantially reduce the amount of data to be transmitted.
Last, there is a method extended from {\itshape private information retrieval} \cite{demmler2018pir}.
Thus, many improvements have been proposed based on these extensions; however, there is still no method to achieve practical efficiency on large scale data in terms of the execution time and communication bandwidth.

Regarding the latter category, there are {\itshape semi-honest} \cite{goldreich2009foundations} or {\itshape malicious} adversaries.
Roughly speaking, a semi-honest adversary is an attacker who tries to infer secret information from the information he obtains, while following correct protocols and not crafting send and receive data, while a malicious adversary is an attacker who crafts send and receive data and executes the protocol as many times as possible to extract secret information.
Generally, the malicious client setting requires a more secure standard and higher costs.
Which model we should secure depends on the applications and situations, but in our scenario, we consider a malicious adversary because it is reasonable to consider that an untrusted server can control and access any computation on the server.
The TEE-based approach can achieve malicious security \cite{subramanyan2017formal}.

We consider the secure-hardware-based approach to be the better option.
Regarding the methodology, we do not have to use the abovementioned cryptographic primitives.
Using Intel SGX, platform verification and transparent memory encryption by the hardware occurs so quickly and thoroughly that a highly efficient PSI can be achieved.
The difference in efficiency is significant, being especially impactful on our choice.
Additionally, the TEE provides a refined security model for a malicious adversary.
\cite{subramanyan2017formal} shows that no operation or inputs expose information regarding the inner state or data of the TEE.
All we need to consider is privacy leakage from outside the TEE and software implementation bugs.

\begin{table}[t]
\centering
\input{tables/psi-feature}
\label{tab:psi-feature}
\end{table}

We present a deeper comparison between the existing cryptography-based PSI and the secure-hardware-based PSI in terms of the efficiency.
We recognize the state-of-the-art approaches as \cite{rindal2017malicious} in terms of the balanced data size setting in the server and client and \cite{chen2017fast} in terms of the unbalanced case.
The latter assumes a semi-honest server and is faster than others in the malicious server model.
Table \ref{tab:psi-feature} shows a comparison of the properties between them and secure hardware (Intel SGX).
It includes a relatively rough estimate of the asymptotic bandwidth and computational costs at every PSI execution.
We denote $n$, $N$ as the client data size and server data size, respectively, assuming $n << N$.
Asymptotic comparisons are acceptable for one purpose because of the large impact of different coefficients.
However, with secure hardware, both communication and computing costs are dramatically more efficient, as they are proportional only to the client data size (Table \ref{tab:psi-feature}).
On the other hand, the secure-TEE-based approach requires RA in advance and a hardware with special functionality on the server side.
Cryptography-based methods do not need any special devices; they consist solely of algorithms.
Rindal et al. \cite{rindal2017malicious} reduce communication cost to $\mathcal(N)$ from the naive $\mathcal(N^2)$ cost by using a variant of Cuckoo hashing in a OT-based method.
In \cite{chen2017fast}, the communication cost is efficiently reduced to $\mathcal{O}(n\log N)$, and the server computational cost is $\mathcal{O}(n)$ homomorphic evaluations on large circuits of size $\mathcal{O}(N/n)$.
These facts show that the secure-hardware-based method demands an extra cost such as special hardware, but is a better choice for large-scale deployment because of the significant efficiency.

\section{Problem Formulation}
\label{sec:3}
We first introduce the trajectory-based PCT scenario, and then we formulate our problem based on the well-studied private proximity testing.

\subsection{Scenario}
\label{sec:3-1}
In our scenario, we assume that trajectory-based PCT is used to prevent the spread of COVID-19.
We consider a centralized architecture that stores the trajectory data of patients with COVID-19 on a central server and accepts PCT requests from users with their trajectory data.
In practice, these patients' trajectories can be received in bulk from public institutions such as a government or health agency.

In the system operation, based on the incubation period of the virus, the server always keeps the trajectory data of the infected patients for the past 14 days \cite{cdcguideline}. 
All the data are periodically updated in batches (e.g., once per day at night), with data being added and deleted.
The server transforms the trajectory data into an appropriate structure in advance and is always ready to accept PCT requests from clients.
A client sends encrypted trajectory data for the past 14 days as a PCT request, and the server performs contact detection and then returns the results to the client.
The results can be time-stamped and signed in SGX as needed so that they can be verified by a third party using an authenticated public key of SGX, allowing clients to use the results for various agencies and events to show that the risk of infection is low.
While client data are protected from any other parties except the TEE by encryption, server-side infected people trajectories can be either confidential or open access ( i.e., the South Korea case \cite{koreaproblem}.
Our system can be applied to both situations; if the infection data should be confidential, a health agency can encrypt the data before uploading the TEE by RA.
Note that the infection data are generally large in size and need to be kept in memory outside of the enclave.
This means that the encrypted data initially uploaded to the TEE are encrypted in the enclave and placed in memory outside of the enclave.
In our experimental implementation, we consider the confidential case which causes additional decryption overheads.


\subsection{Problem Statement}
\label{sec:3-2}

\noindent
{\bfseries Trajectory-based PCT.}
The trajectory-based PCT protocol is an asymmetric protocol between a client and a server.
When a client wants to know the contact with trajectories stored on a server, this protocol returns 1 or 0 to the client depending on the result, and does not disclose the private information of the client to the server.
In the use case for infections, each client has a set of trajectory data for one person, and the server has trajectory data for many infected patients.

In conventional private proximity testing \cite{narayanan2011location}, when two people, user $u$ and $v$, have geographic data $X_i$ that consist of location $l^{(i)}_t$ (= (latitude, longitude), e.g., (40.74836, -73.98562)) of time $t$ ($i=u, v$), user $u$ executes the protocol and obtains the following result:
$$
\begin{cases}
    \;1 \;\; (\,|| l^{(u)}_t - l^{(v)}_t || \le \Theta\,) \\
    \;0 \;\; (\, others \,)
\end{cases}
$$

\noindent
where $\Theta$ is a proximity threshold.
After that, $v$ does not learn any information about $X_u$ and $u$ does not learn information except whether $|| l^{(u)}_t - l^{(v)}_t || \le \Theta $.

In a similar vein, trajectory-based PCT can be represented as an extension of this formulation.
For contact tracings, a contact can be determined according to human time-series tracking data.
We can perform private proximity testing by extending single geographic data to time-series trajectory data.
The threshold can also be extended to 2D thresholds to check the spatiotemporal proximity.
PCT allows capturing indirect contacts by examining whether patients are in the proximate place within a specific time period.
Therefore, we obtain the following formula denoting the trajectory data of user $i$ as $X_i=(x^{(i)}_1=(t^{(i)}_1,l^{(i)}_1),...,x^{(i)}_n=(t^{(i)}_n,l^{(i)}_n))$ (e.g., $x_{100} = $(2020/12/20 12:00:00, 40.74836, -73.98562)), with which the result of contact between $u$ and $v$ can be obtained:
\begin{eqnarray}
\label{eq:1}
\begin{cases}
    \;1 \;\; (\exists \,x^{(u)}_i \in X_{u}\,, \;x^{(v)}_j \in X_{v} \,s.t.\\
    \;\;\;\;\;\;\;\; ||l^{(u)}_i - l^{(v)}_j|| \le \Theta_{geo} \,\; \mathrm{and} \,\; ||t^{(u)}_i - t^{(v)}_j|| \le \Theta_{time} \,) \\
    \;0 \;\; (\, others \,)
\end{cases}
\end{eqnarray}

\noindent
where $\Theta_{geo}$ and $\Theta_{time}$ are the spatial and temporal proximity thresholds, respectively. 
Furthermore, $v$ does not learn any information about $X_u$, and $u$ can obtain only 1 or 0 regarding $X_v$ in this protocol.
The thresholds are generally given by medical and epidemiological experts and may be updated.
We define this procedure as {\itshape trajectory-based PCT}, which can capture indirect contacts by having a certain width in the time direction.
Moreover, we can extend the definition so that the duration of exposure to an infected user can be considered, which is recognized as an important factor of COVID-19 transmission \cite{cevik2020sars, cdcguideline}.
Let $\Theta_{doe}$ be the risky duration of exposure, the following formula is obtained:
\begin{eqnarray}
\label{eq:2}
\begin{cases}
    \;1 \;\; (\;\exists \tau \;\;\; s.t. \;\;\; \forall i= (\tau, \tau+1,...,\tau+\Theta_{doe}) \\
    \;\;\;\;\;\;\;\;  \exists \,x^{(u)}_i \in X_{u}\,, \;x^{(v)}_j \in X_{v}, \;\;\; s.t. \;\;\; ||l^{(u)}_i - l^{(v)}_j|| \le \Theta_{geo} \,\; \mathrm{and} \,\; ||t^{(u)}_i - t^{(v)}_j|| \le \Theta_{time} \,) \\
    \;0 \;\; (\, others \,)
\end{cases}
\end{eqnarray}

\begin{figure}[t]
    \centering
  \includegraphics[height=200pt]{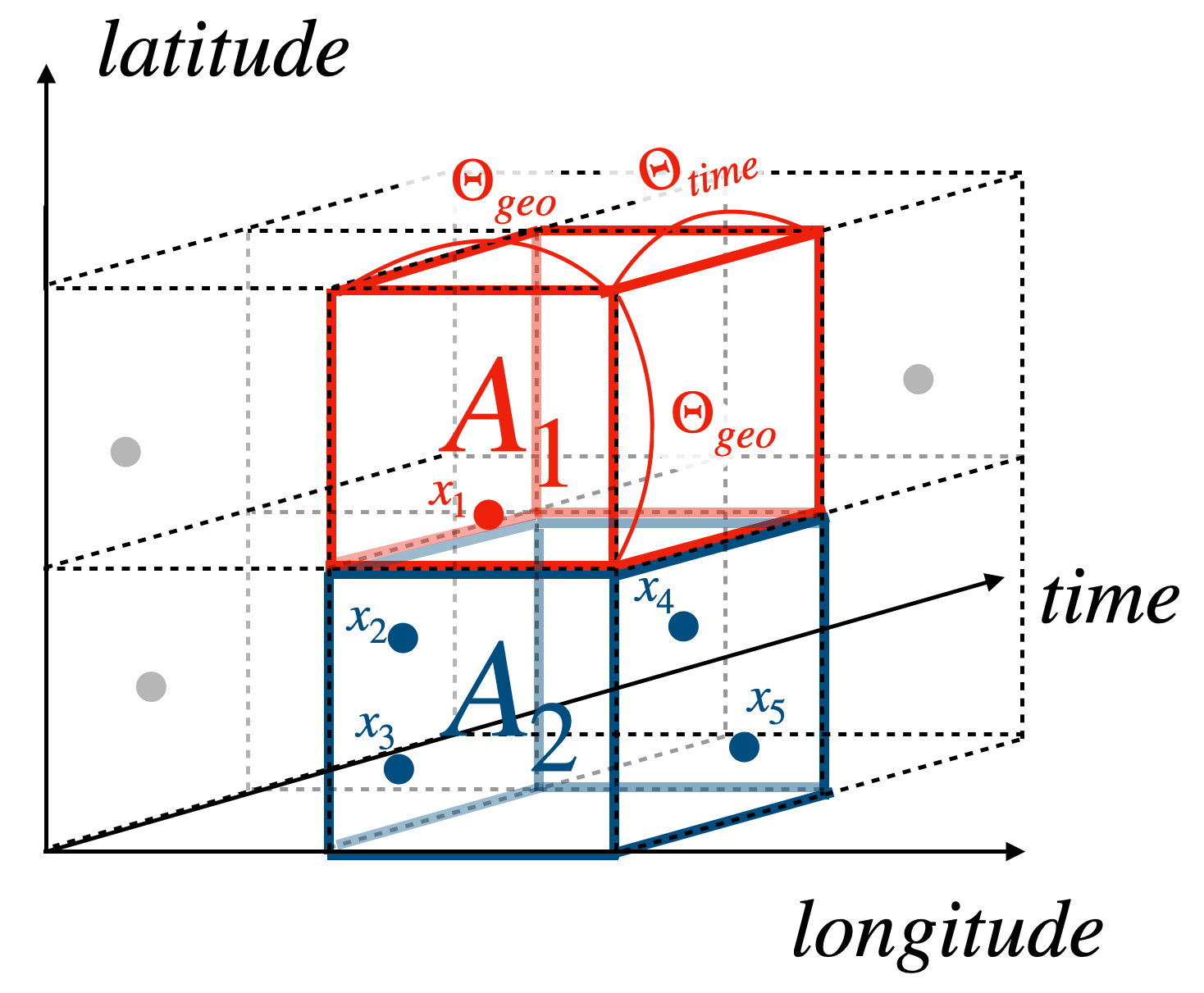}
  \caption{Spatiotemporal Private Set Intersection.}
  \label{fig:psi_def}
\end{figure}

Finding the exact solution of Eq.(\ref{eq:1}) and Eq.(\ref{eq:2}) for the sampled discrete trajectory data is computationally inefficient.
Therefore, we simplify this problem by mapping continuous space to discrete space for computational efficiency, where we approximate the PCT problem to the PSI problem.
We denote $\mathbb{A}$ as the set of all symbols in a discrete space and $A_{i}\in\mathbb{A}$ as the $i$-th element.
By mapping $f_{\Theta}: x \rightarrow A$, we can map any point $x$ in the trajectory data to a single symbol $A$.
We call this mapping an ``encoding" and introduce the corresponding method in Section \ref{sec:5}.
The encoding, $f_{\Theta}$, must be adjustable according to the granularity parameters $\Theta_{geo}$ and $\Theta_{time}$, which correspond to the size of the predefined subspace in the 3D spatiotemporal space, and each subspace corresponds to one unique symbol, as shown in Figure \ref{fig:psi_def}.
For example, suppose $f_{\Theta}(x_1)=A_{1}$, $f_{\Theta}(x_2)=A_{2}$, $f_{\Theta}(x_3)=A_{2}$, $f_{\Theta}(x_4)=A_{2}$, $f_{\Theta}(x_5)=A_{2}$, trajectory point $x_1$ is mapped to $A_{1}$, and $x_2$,$x_3$,$x_4$, and $x_5$ are mapped to $A_{2}$ in Figure \ref{fig:psi_def}.
We determine contact by considering the intersection of these symbol sets between $u$ and $v$.
This can be formulated as follows:
\begin{eqnarray}
\label{eq:3}
\begin{cases}
    \;1 \;\; \bigl(\exists \, \mathbf{A}^{(u)} \cap \mathbf{A}^{(v)} \neq \emptyset \,\,\,\,s.t.\\
    \;\;\;\;\;\;\;\; \mathbf{A}^{(u)} = \{\, f_{\Theta}(x^{(u)}_i)\;| \;\;x^{(u)}_i \in X_{u} \,\} \,\; \mathrm{and} \,\; \mathbf{A}^{(v)} = \{\, f_{\Theta}(x^{(v)}_j)\;|\;\; x^{(v)}_j \in X_{v} \,\} \,\bigl) \\
    \;0 \;\; (\, others \,)
\end{cases}
\end{eqnarray}
\noindent
$v$ does not learn any information about $X_u$, and $u$ can obtain only 1 or 0 regarding $X_v$ in this problem.
We name this problem the {\itshape spatiotemporal PSI} or {\itshape stPSI} for short.
Considering the duration of exposure and given threshold parameters $\Theta=(\Theta_{geo}, \Theta_{time})$ and $\Theta_{doe}$, the problem can be formulated as follows:
\begin{eqnarray}
\label{eq:4}
\begin{cases}
    \;1 \;\; \bigl(\;\exists \tau \;\;\; s.t. \;\;\; \forall i= (\tau, \tau+1,...,\tau+\Theta_{doe}) \\
    \;\;\;\;\;\;\;\; (\exists \, \mathbf{A}^{(u)} \cap \mathbf{A}^{(v)} \neq \emptyset \,\,\,\,s.t.\\
    \;\;\;\;\;\;\;\; \mathbf{A}^{(u)} = \{\, f_{\Theta}(x^{(u)}_i)\;| \;\;x^{(u)}_i \in X_{u} \,\} \,\; \mathrm{and} \,\; \mathbf{A}^{(v)} = \{\, f_{\Theta}(x^{(v)}_j)\;|\;\; x^{(v)}_j \in X_{v} \,\}) \,\bigl) \\
    \;0 \;\; (\, others \,)
\end{cases}
\end{eqnarray}
Note that the worst computational complexity this problem causes is the same as in Eq.(3) since we need to sequentially check the $u$'s of all the data only once, both of which are checked $\mathcal{O}(|X_u|)$ times.




\noindent

In this work, we consider the stPSI as a contact between $u$ and $v$.
Hence, we can reduce the problem to one that is simple enough to be computed with existing confidential computing stacks in a practical computational time.
However, the problem is basically an approximation of trajectory-based PCT.
Given the threshold parameter $\Theta$, this approximation causes some detection errors.
We carefully analyze these errors in Section \ref{sec:5-5} and show that there is a trade-off between false positives and false negatives due to the proposed algorithms and thresholds.

Additionally, we describe the important requirements of trajectory-based PCT.

\noindent
{\bfseries Efficiency.}
Trajectory-based PCT requires efficiency in several aspects.
\begin{itemize}
    \item The first is the response throughput since the server will always be exposed to requests from a large number of users. It can be a substantial workload in this centralized protocol.
    \item The second aspect is the bandwidth. Since the protocol is applied to many users, it is necessary to reduce the bandwidth for communication efficiency.
    \item The third aspect is scalability. For instance, for COVID-19, the size of the infected patient data and the user data may increase in the event of infection spreading.
\end{itemize}
The efficiency requirements depend entirely on the context in which PCT is deployed and are determined by the amount of users, frequency of use, number of data, etc.

\noindent
{\bfseries Security.}
Security concerns in general PCT systems are manifold.
The concerns must include violating the privacy of participants by revealing personal information such as location, inducing errors in PCT results by introducing false information, and violating the integrity and availability of the system.
In our scenario, we believe it is reasonable to assume a malicious client, a malicious server, and a semi-honest health agency as the attacker models. Because any client can participate in a service, and because it is not obvious how the server runs in a remote environment, we should assume that it has a full control over the operating system and/or hypervisor, memory hardware units, and packet flow in the network and uses them to attack the system. In addition, the selected health agency is publicly authenticated. 
In the centralized architecture used in our scenario, some of the attacks that are often of concern in BLE-based PCT methods are not possible (e.g., carryover attack \cite{becker2019tracking}, paparazzi attack \cite{vaudenay2020analysis}, Orwell attack \cite{avitabile2020towards}, etc.).
On the other hand, we should consider the following attacks;
\begin{itemize}
    \item Denial-of-service (DoS) attacks: malicious clients send many requests to the server to bring down the system.
    \item Query-abusing attacks: privacy violation attacks are performed to obtain an infected person's data by a malicious client's queries.
    \item Side-channel attacks: the malicious server causes information leakage by side-channel attacks on communication paths and within the server.
    \item False answering: the malicious server answers clients by sending fake results.
    \item Fake data injection: the malicious server injects fake infected data into the system.
    \item Replay: someone catches the client request information by communication interception and reuses it to obtain the client's PCT result.
\end{itemize}
Our proposed trajectory-based PCT system should have countermeasures to prevent or mitigate all of these attacks.

\noindent
{\bfseries Flexibility.}
The rules of risky contact in the case of COVID-19 have been updated as the understanding of the virus has improved \cite{kgwreport} as we explained Section \ref{sec:1}.
In the trajectory-based PCT, flexibility requires that $(\Theta_{geo}, \Theta_{time}, \Theta_{doe})$ be parameterized and changeable in the PCT system.
For example, these parameters need to be changed to minimal values if it is found after the system is released that only direct contact needs to be captured because of the virus's capacity for transmission.
However, we do not believe that these parameters need to be parameterized at the client query level (i.e., clients can choose the parameters).
In other words, we assume that there is one global condition that is epidemiologically determined to be important for infection prevention.
In addition, if the client has control over the parameters, it can expose information about the infected person's data, creating an unnecessary security risk.

We believe that these parameters should be updated only once a day at most.
Therefore, it is reasonable to apply the changes in these parameters to the data at the same time as the batch update of the infected data on the server, which justifies our proposed method.
In our proposed method, we encode the data in advance with a granularity that matches the parameters and then use the PSI to detect contact decisions.

\noindent
{\bfseries Accuracy.}
Accuracy requires to achieve the high correct detection rate.
Since virus infection information is very sensitive, it is necessary to reduce the percentage of responses to clients that contain false positives or false negatives.
In addition, when a response is \textit{false}, what kind of \textit{false} it is, i.e., whether it is nearly correct or randomly generated, has completely different meanings in terms of whether there is an upper limit to the error or not.
For example, even if the epidemiological false positive is somewhat large, it may be effective in preventing infection depending on the phase of infection \cite{pandl2021detection}.

There are several aspects of trajectory-based PCT that can cause a \textit{false} outcome.
The first one is the error of trajectory data collection.
Many studies have shown that the accuracy of GPS-based trajectory data collection is improved \cite{ali2018senseio, huang2018cts}; however, the accuracy is degraded especially in indoor environments.
Also, if they stay somewhere for too short a time, they may not be captured when the trajectory data is collected.
On the other hand, it is possible to improve the collection accuracy by combining wireless devices such as indoor Wi-Fi \cite{kulshrestha2017smartits}.
Our work assumes that these state-of-the-art devices collect highly accurate trajectory data, and this false is beyond the scope.
The second is a \textit{false} result caused by the stPSI approximation.
It is necessary to clarify how many false positives and false negatives are generated by the approximation and what is the upper limit of the error.
Through extensive experiments, we empirically clarify how much error is caused by the stPSI.
The third is the \textit{false} result caused by the physical environment, which cannot be captured by location information alone.
For example, if a person is riding in a car or spending time in an adjacent room, a false positive may occur if only trajectory data are used.
This error is difficult to determine from the trajectory data and requires a different type of data; we consider it to be the topic of future work.

\section{System Overview}
\label{sec:4}
We introduce an overview of the system.
Table \ref{tab:params} shows the symbols and parameters that are used in the rest of the paper.

\begin{table}[t]
\centering
\input{tables/parameters}
\caption{Symbols and parameters.}
\label{tab:params}
\end{table}

\begin{figure}[t]
  \includegraphics[width=\linewidth]{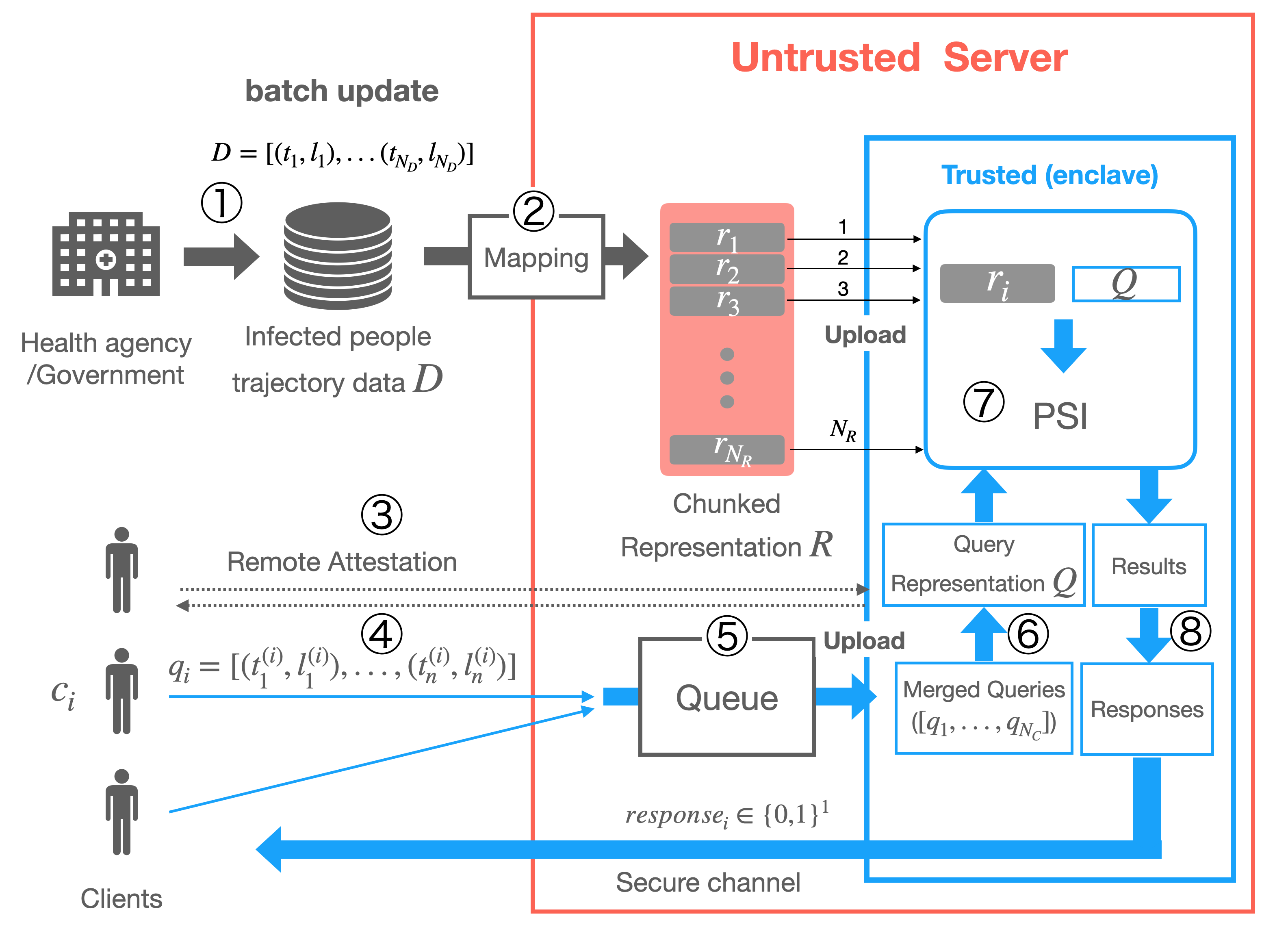}
  \caption{Architecture overview: circled numbers correspond to the steps shown in {\itshape System Overview}.}
  \label{fig:carousel}
\end{figure}

Figure \ref{fig:carousel} shows an overview of our architecture with a trusted enclave.
Our method consists of several steps, including the transformation of data maintained on the server side and the transformation of data sent from the clients, as follows. 

First, we describe the data of infected patients on the server.

\noindent
\circled{1} : (Update master data) The health agency updates the infected patient data $D$ in batch processing.
$D$ is in the raw form of trajectory data $D=[(t_1,l_1),...,(t_{n_R},l_{n_R})]$ and does not have to include the user IDs since there is no need to distinguish trajectory data by infected users.

\noindent
\circled{2} : (Mapping) Step 2 is executed in the same batch processing as step 1.
We map from the raw data format $D$ to efficient dictionary representation $R$ with the function ${\textsc{mapToChunkedDictionary}}$.
This mapping function includes encoding, chunking, and transforming into the dictionary representation.
Encoding is to encode each piece of trajectory data into 1D bytes of representation. 
It corresponds to $f_{\Theta}$ in Eq. (\ref{eq:3}) (\ref{eq:4}).
Chunking is to split the dataset into $N_R$ chunks.
Transforming is to transform each chunk into a dictionary representation and into $R$ that consists of $N_R$ chunks $r_i$ $(i=1,...,N_R)$, where each chunk fits in the enclave memory limitation.
How to represent the chunked data specialized in the PSI under the SGX memory constraint is our challenge.
These encoding and compression schemes are described in Section \ref{sec:5}.
In the case where the infected data are to be kept secret, the encrypted infected data are uploaded to SGX by the health agency, step 2 is performed in SGX, and the binary data of the encrypted $R$ are placed in the memory of the untrusted server.
The key used for encryption during the upload is obtained from SGX by the health agency through RA.

The next part is the processing of queries from clients.

\noindent
\circled{3} : (Remote attestation) The client verifies the remote enclave through the RA protocol before sending the request to the server.
The client can confirm that the enclave has not been tampered with and then securely exchange the key with the enclave.
After that, the shared key are used to encrypt the data, which enables secret communication to the remote enclave through a secure channel.
Note that a secure channel is a communication channel where only the query data in the request are encrypted and kept secret.
Client metadata (e.g., IP address) exchanged in the application layer is not kept secret from the server side and need to be used in the response.
The response data are encrypted and kept secret, as well as the request query data.

\noindent
\circled{4} : (Request) Many clients send PCT requests to the server.
In the figure, $c_i$ sends $q_i$ as a request parameter that contains 14 days of her trajectory data.
Trajectory data are encoded by $f_{\Theta}$ before encryption; thus, the server and client share the parameter $\Theta$ in advance.
$q_i$ is encrypted in all the untrusted areas after leaving the client environment and is visible only in the verified enclave.

\noindent
\circled{5} : (Queuing) Until a certain number ($N_c$) of requests are accumulated, $q_i$ is queued outside the enclave, and they are passed to the enclave together by the {\textsc{loadToEnclave}} function.
This function is actually implemented by the so-called {\itshape ECALL} to invoke an SGX function.
We aim to optimize the loading process for multiple (e.g. 1000) users by batch processing, mitigating the overheads invoked.

\noindent
\circled{6} : (Mapping) After uploaded to the trusted enclave, the data are finally decrypted.
Inside the enclave, all $q_i$ are grouped together and mapped to query representation $Q$ using {\textsc{mapToArray}}.
While these query data are private and cannot be handled outside the enclave, the size of the enclave memory is strictly limited.
Therefore, encoding trajectory data to small bytes is critical.

\noindent
\circled{7} : (Contact detection) The chunked data $r_i$ are imported into the enclave one by one, and we compute the set intersection of $r_i$ and $Q$ in the enclave.
This can be done by checking the bytes-based match.
In the result, only queries with a true result for a set intersection are recorded as positive.
If a query is found to be positive, we can reduce some overhead by not computing the PSI for subsequent $r_i$ for the query.

\noindent
\circled{8} : (Response construction) After the iterations for all the chunks are completed, responses for all clients are constructed from the results and complete query data $q_i$ $(i=1,...,N_C)$ inside the trusted enclave by {\textsc{constructResponses}}.
This process can be carried out by simply encrypting the results (positive or negative) for each client inside the enclave.
Finally, the process returns the encrypted result through the secure channel to each client.

\section{Spatiotemporal PSI}
\subsection{Trajectory Data Representation}
\label{sec:5}
In this subsection we focus on the trajectory data representation, which is optimized for PSI processing in the memory constraint of Intel SGX.
The most important issue is how to represent each trajectory data point.
This is called encoding and corresponds to $f_{\Theta}$ in Eq. (\ref{eq:3}) (\ref{eq:4}).
We need to encode different trajectory data into unique 1D data to solve the trajectory-based PCT as the stPSI, as described in Section \ref{sec:3-2}.
In addition, we need to make the encoded data as small as possible.
The compact representation contributes to the whole system performance, and the compressibility contributes to the performance of the PSI part which is the core component of our system.

Additionally, we carefully develop the dictionary representation $R$ ($=(r_1,...r_{N_R})$) obtained by the mapping in step 2.
$R$ should satisfy the following constraints: First, a memory-efficient data structure storing trajectory data should be used to overcome the severe memory constraints of SGX. Second, a fast search should be implemented for a fast PSI. Third, a deterministic search method for an accurate PSI should be provided.
The standard dictionary representations do not fulfill these requirements.
We consider Hashmap as a baseline.
Hashmap ideally supports the $\mathcal{O}(1)$ key-based search.
While it provides desirable search performance and allows a deterministic search, it fails to satisfy the efficiency requirements because its size increases linearly with the size of the data.
A smaller data structure is preferable in our setting because the overheads caused by SGX are considerably heavy.
While probabilistic data structures such as the Bloom filter provide the same search speed performance as that of Hashmap and superior memory efficiency, they do not satisfy deterministic search requirement.
It causes random and unpredicted false positives.

Our proposed method to achieve the desired dictionary representation is a combination of effective trajectory data encoding and storing into a fast succinct trie (FST) \cite{zhang2018surf}.
An FST is a data structure proposed in \cite{zhang2018surf} and is the base of Succinct Range Filter (SuRF).
The SuRF can improve efficiency for queries such as match and range in exchange for false positives, while the FST does not allow any false positives.
The FST has basically the same properties as those of trie, but its internal representation is closer to being succinct, and it has a particular strength in spatial efficiency.
For more information on the FST, please refer to \cite{zhang2018surf}.
Similar to the basic trie, the FST provides high compression performance for highly similar data by sharing a common prefix of encoded bytes in a single node.
Essentially, our aim is to ensure that the encoding process transforms trajectory data into highly similar byte sequence representations and then utilizes the similarity to create a compressed dictionary representation using an FST.

\begin{figure}[t]
  \includegraphics[width=0.9\linewidth]{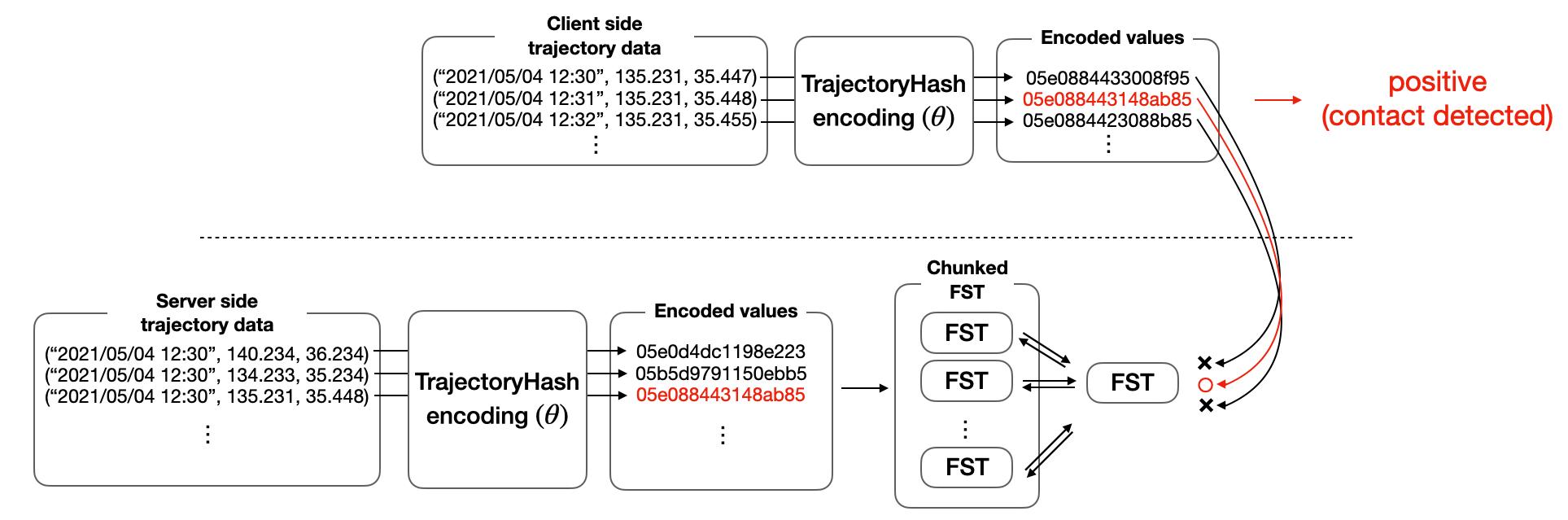}
  \caption{An overview of running example from raw trajectory to the final risk assessment.}
  \label{fig:run_ex}
\end{figure}

Figure \ref{fig:run_ex} shows an overview of running example from raw trajectory to the final risk assessment.

\subsubsection{TrajectoryHash}
\label{sec:5-1}
We introduce encoding, which corresponds to $f_{\Theta}$ in the stPSI as described in Section \ref{sec:3-2}.
The encoding should satisfy the following 3 properties.
First, there is an injective function between different trajectory data and unique byte sequences.
Obviously, if this property is not satisfied, the PSI cannot be performed correctly.
Second, the encoded value should be small in size.
The space after mapping should be as small as possible because if it is small, all data, including both the server data and the queries from the client, will be small.
This is the ideal situation for TEE-based secure computation.
Our previous work \cite{kato2020secure} lacks this aspect; string-level merging caused the binary size of the encoded values to grow.
The other desired property is that the encoded values should have many similarities because we can efficiently store within the FST.
We introduce {\itshape TrajectoryHash} encoding and show that it satisfies all properties.

The trajectory data $X$ consist of an array of tuples of temporal data and geographical data, such as the UNIX epoch and tuples of latitude and longitude, as follows.

\begin{equation*}
\begin{gathered}
\mathbf{X} = [x_1=(t_1,l_1),...,x_n=(t_n, l_n)] \\
t_k \in \mbox{time (UNIX epoch)} \\
l_k=(l_{k, lat}, l_{k, lon}) \in \mbox{coordinate ((latitude,\,longitude)})
\end{gathered}
\end{equation*}

\begin{figure}[t]
  \begin{algorithm}[H]
    \caption{{\textsc{TrajectoryHash}}}
    \label{alg:th}
    \begin{algorithmic}[1]
      \Require $t,$, $(l_{lat}, l_{lng})$, $\theta_{time}$, $\theta_{geo}$, $t_{start}$, $t_{end}$
      \Ensure $hash$
     \State $b_1, b_2 \leftarrow {\textsc{QuadKeyEncode}}(l_{lat}, l_{lng}, \theta_{geo})$ \algorithmiccomment{$b_1$, $b_2$ are binary sequence with $\theta_{geo}$ length}
      \State $b_3 \leftarrow {\textsc{PeriodicEncode}}(t, \theta_{time})$
     \algorithmiccomment{$b_3$ are binary sequence}
      \State $binary \leftarrow {\textsc{BitMix}}(b_1, b_2, b_3)$
      \State $hash \leftarrow {\textsc{ByteEncode}}(binary)$
      \State \Return $hash$
    \Procedure{\textsc{QuadKeyEncode}}{$l_{lat}, l_{lng}, \theta_{geo}$}
        \State $maxLatitude \leftarrow 360\arctan (\exp(\pi)) / \pi - 90$  \algorithmiccomment{$\simeq 85.05112877980659$}
        \State $l_{lat} \leftarrow \min(maxLatitude, \max(-maxLatitude, l_{lat}))$ \algorithmiccomment{for clipping}
        \State $px \leftarrow \cfrac{l_{lng}+180}{360}$ \algorithmiccomment{transform to the Tile Coordinates}
        \State $py \leftarrow \Bigl(\cfrac{1}{2} - \cfrac{1}{\pi}\log{\cfrac{1+\sin{(l_{lat}\times \pi/180)}}{1-\sin{(l_{lat}\times \pi/180})}}\Bigl)$
        \State $mapSize \leftarrow 2^{\theta_{geo}}$ \algorithmiccomment{map consists of $2^{\theta_{geo}}\times 2^{\theta_{geo}}$ areas}
        \State $x \leftarrow {\textsc{Floor}}(px\times mapSize)$ \algorithmiccomment{round down function}
        \State $y \leftarrow {\textsc{Floor}}(py\times mapSize)$
        \State $Xbinary \leftarrow {\textsc{AsBinary}}(x)$ \algorithmiccomment{get as bit array representation}
        \State $Xbinary \leftarrow {\textsc{ZeroPadding}}(Xbinary, \theta_{geo})$ \algorithmiccomment{padding 0 to $\theta_{geo}$ length}
        \State $Ybinary \leftarrow {\textsc{AsBinary}}(y)$
        \State $Ybinary \leftarrow {\textsc{ZeroPadding}}(Xbinary, \theta_{geo})$
        \State \Return $Xbinary, Ybinary$
    \EndProcedure
    \Procedure{\textsc{PeriodicEncode}}{$t, t_{start}, t_{end}, \theta_{time}$}
        \State $maxLength \leftarrow {\textsc{Length}}({\textsc{AsBinary}}(t_{end} - t_{start}))$ \\ \algorithmiccomment{maximum bit length to represent the period $t_{start}$ to $t_{end}$}
        \State $t_{diff} \leftarrow t - t_{start} $
        \State $shift \leftarrow 32 - \theta_{time}$ \algorithmiccomment{32 = max bit length of UNIX epoch}
        \State $t_{diff} \leftarrow {\textsc{Floor}}(t_{diff} / 2^{shift})$ \algorithmiccomment{right shift $t_{diff} >> shift$}
        \State $binary \leftarrow {\textsc{AsBinary}}(t_{diff})$
        \State $binary \leftarrow {\textsc{ZeroPadding}}(binary, maxLength-shift)$
        \State \Return $binary$
    \EndProcedure
    \end{algorithmic}
  \end{algorithm}
\end{figure}

\noindent
$t_1$ and $t_n$ are determined as $t_{start}$ and $t_{end}$ considering conditions such as the lifespan of the virus. (e.g., in the case of COVID-19, we currently believe it is 14 days.)
Algorithm \ref{alg:th} shows the pseudocode of {\textsc{TrajectoryHash}}.
The input parameters $\theta_{geo}$ and $\theta_{time}$ correspond to $\Theta_{geo}$ and $\Theta_{time}$, respectively.
While $\Theta_{geo}$ and $\Theta_{time}$ directly express the spatial distance as in Figure \ref{fig:psi_def}, note that $\theta_{geo}$ and $\theta_{time}$ express granularity on a different scale, as shown in Table \ref{tab:paramthetageo}, \ref{tab:paramthetatime}.
The algorithm of the encoding is based on two encodings {\textsc{QuadKeyEncode}}, {\textsc{PeriodicEncode}} and a binary-level mixing function, {\textsc{BitMix}}.
ST-Hash \cite{guan2017st} is similar to our encoding.
The part to be mixed at the binary level is the same, but the 2 encoding methods and the motivation are different.
We use {\textsc{QuadKeyEncode}} and {\textsc{PeriodicEncode}} to preserve the trajectory data similarity and hierarchical structure to compress the trajectories.

\begin{table}[t]
 \begin{minipage}[t]{0.45\hsize}
  \begin{center}
    \input{tables/param_theta_geo}
  \end{center}
    \caption{Approximate scale of the parameter $\theta_{geo}$.}
    \label{tab:paramthetageo}
 \end{minipage}
 \hspace{0.1cm}
 \begin{minipage}[t]{0.45\hsize}
  \begin{center}
    \input{tables/param_theta_time}

  \end{center}
    \caption{Approximate scale of the parameter $\theta_{time}$.}
    \label{tab:paramthetatime}
 \end{minipage}
\end{table}

\begin{figure}[t]
  \includegraphics[width=0.9\linewidth]{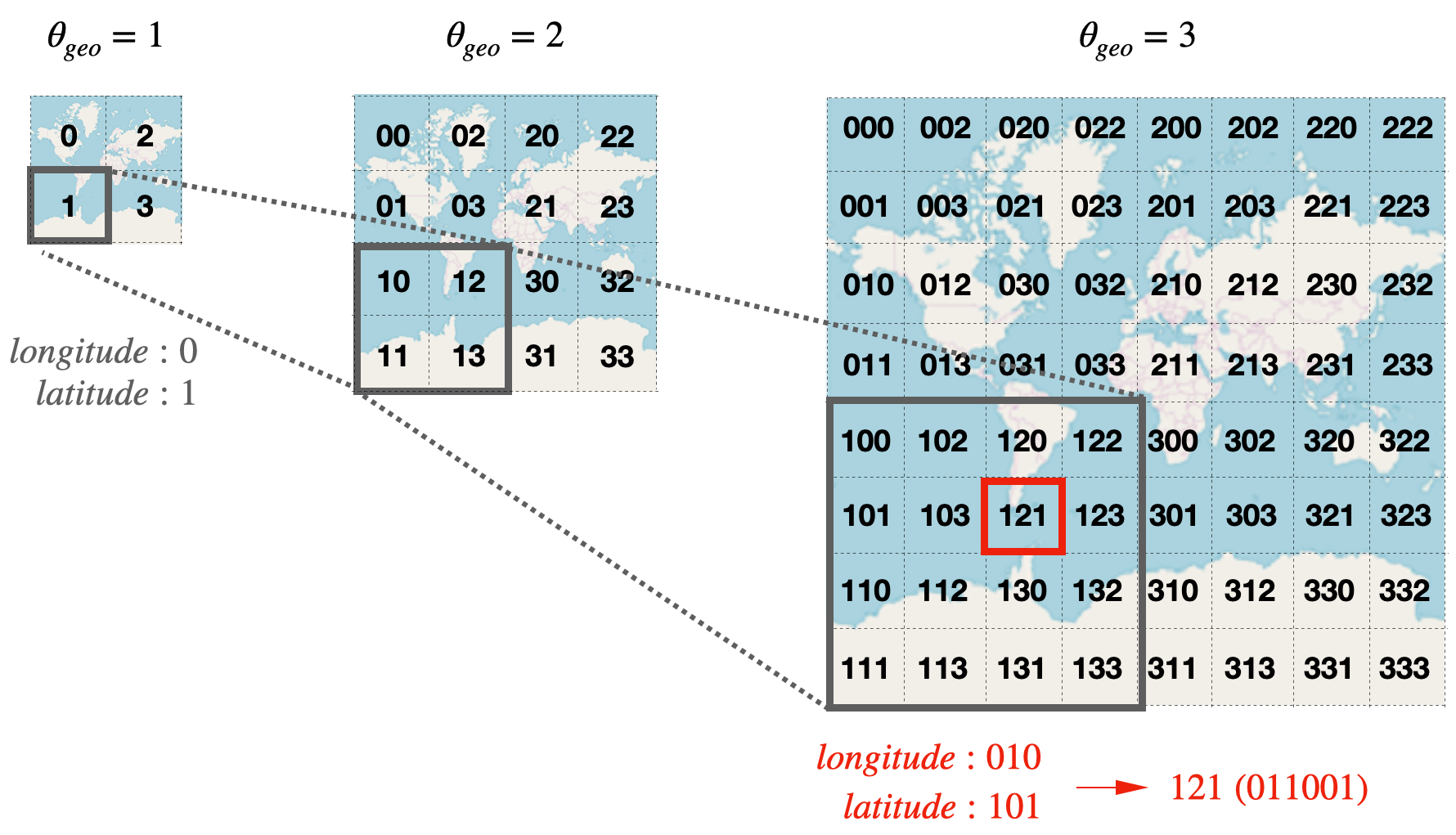}
  \caption{quadkey and {\textsc{QuadKeyEncode}}.}
  \label{fig:quadkey}
\end{figure}

{\textsc{QuadKeyEncode}} is based on the quadkey introduced by Bing Map \cite{quadkey}, which is a method of encoding into bits in the {\itshape tile coordinate} space, recursively dividing into two parts according to a given level, as shown in Figure \ref{fig:quadkey}.
Note that in our method, {\textsc{QuadKeyEncode}} outputs separated binaries. 
As we can see in the figure, while we obtain "212"(=$100110$ in binary) using quadkey encoding, {\textsc{QuadKeyEncode}} outputs $101$ and $010$.
This algorithm is described in detail in Algorithm \ref{alg:th}.
The parameter $\theta_{geo}$ and the approximated distance included in the square in tile coordinates are shown in Table \ref{tab:paramthetageo}. 
Strictly speaking, it is correct for both the latitude and longitude at the equator, but the tile length of the latitude is slightly variable according to the height of the latitude, e.g., in New York, $\theta_{time}=22$ corresponds to $\Theta_{time}=1.83$.
Using this encoding, we obtain unique binaries for each distinguishable area by $\theta_{geo}$.
Moreover, we can keep the hierarchical structure and similarity of trajectory locations in the binary representation.
For instance, given $\theta_{geo}=16$, $(l_{lng}, l_{lat}) = (135.3214557, 30.4564223)$, we obtain the output 1110000000111010 and 0110100100111110 as binaries.

{\textsc{PeriodicEncode}} is optimized to discretize the time data over a specific given period and at specific given time intervals.
This encoding outputs bits with minimum length that can express a distinct time interval according to given $\theta_{time}$ in the period $t_{start}$ to $t_{end}$.
Given 14 days as the period, the relation between parameter $\theta_{time}$ and the approximate time interval is as shown in Table \ref{tab:paramthetatime}.
The final output length is determined by both $\theta_{time}$ and ($t_{start}$, $t_{end}$).
For example, given
$(t, t_{start}, t_{end}, \theta_{time}) = (``2020/10/10\; 10\colon00",\,``2020/10/05\; 00\colon00",\,``2020/10/19\; 00\colon00",24)$
, the processing is carried out in detail as follows:
\begin{align*}
&t_{end} - t_{start} = 1603065600 - 1601856000 = 1209600 \\
&maxLength = 21 \;\; (1209600 < 2^{21}=2097152) \mbox{    (line 21)} \\
&t_{diff} = 1602324000 - 1601856000 = 468000 \mbox{    (line 23)} \\
&shift = 32 - 24 = 8 \mbox{    (line 24)} \\
&t_{diff} = 468000 / 2**8 = 1828 \mbox{    (line 25)}\\
&binary = {\textsc{AsBinary}}(1828) = 11100100100 \mbox{    (line 26)} \\
&binary = {\textsc{ZeroPadding}}(11100100100, 21-8) = 0011100100100 \mbox{    (line 27)} \\
\end{align*}
Finally, we obtain 0011100100100 as a binary.
In this way, we obtain the minimum representation to express trajectory time information and preserve the time representation similarity of the trajectories in the period while adjusting intervals to the given granularity parameter $\theta_{time}$.
Obviously, this gives an information-theoretic lower bound on the byte size, since it is the smallest bit representation that identifies a given period.
Furthermore, in terms of location information, since the bit array obtained from {\textsc{QuadKeyEncode}} is also the size of the information-theoretic lower bound for identifying individual regions in the map at the given scale, TrajectoryHash is a succinct representation of the trajectory data for a given granularity.

\begin{figure}[t]
 \begin{minipage}[b]{0.4\hsize}
    \begin{center}
      \includegraphics[width=\linewidth]{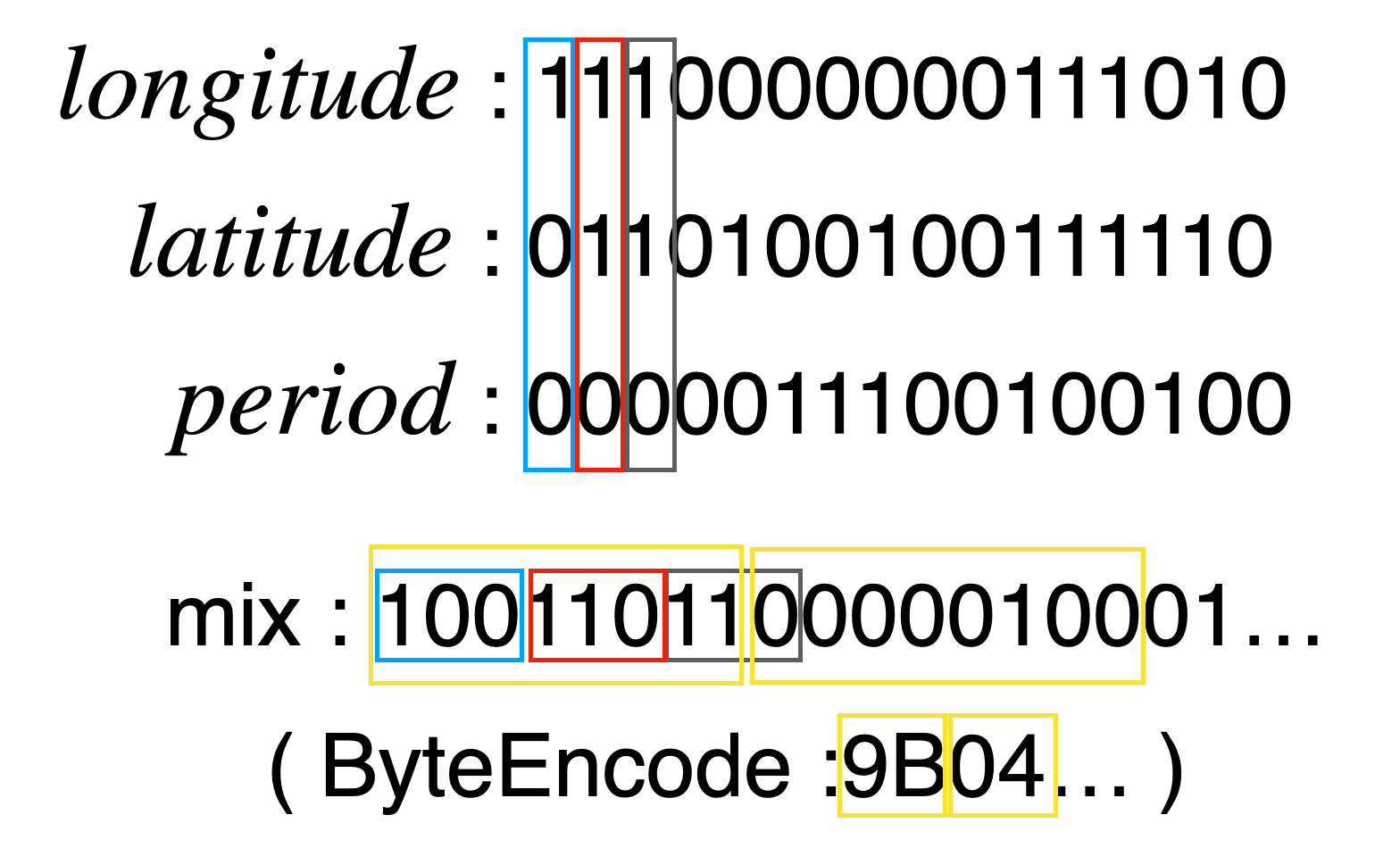}
      \caption{Mixing TrajectoryHash.}
      \label{fig:mixTH}
    \end{center}
 \end{minipage}
 \begin{minipage}[b]{0.5\hsize}
    \begin{center}
      \includegraphics[width=\linewidth]{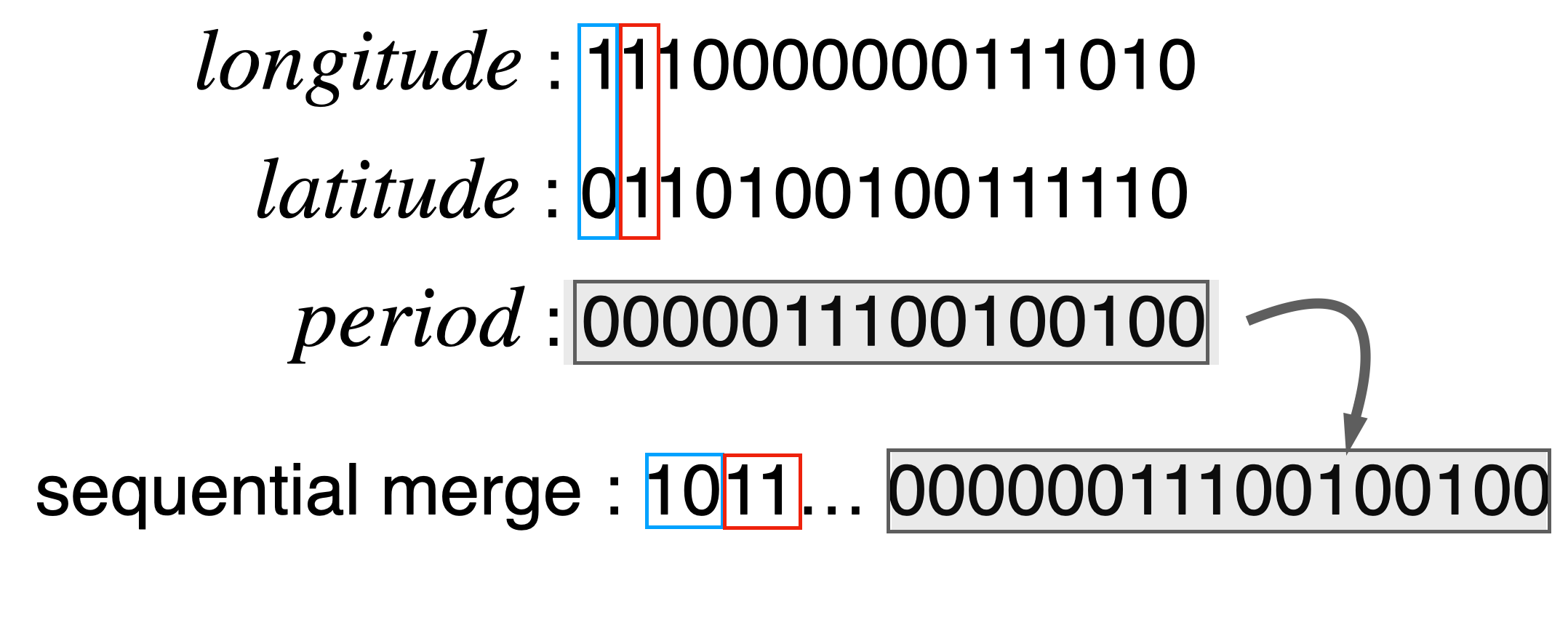}
      \caption{Sequential merge TrajectoryHash.}
      \label{fig:seqTH}
    \end{center}
 \end{minipage}
\end{figure}

\begin{figure}[t]
    \includegraphics[height=120pt]{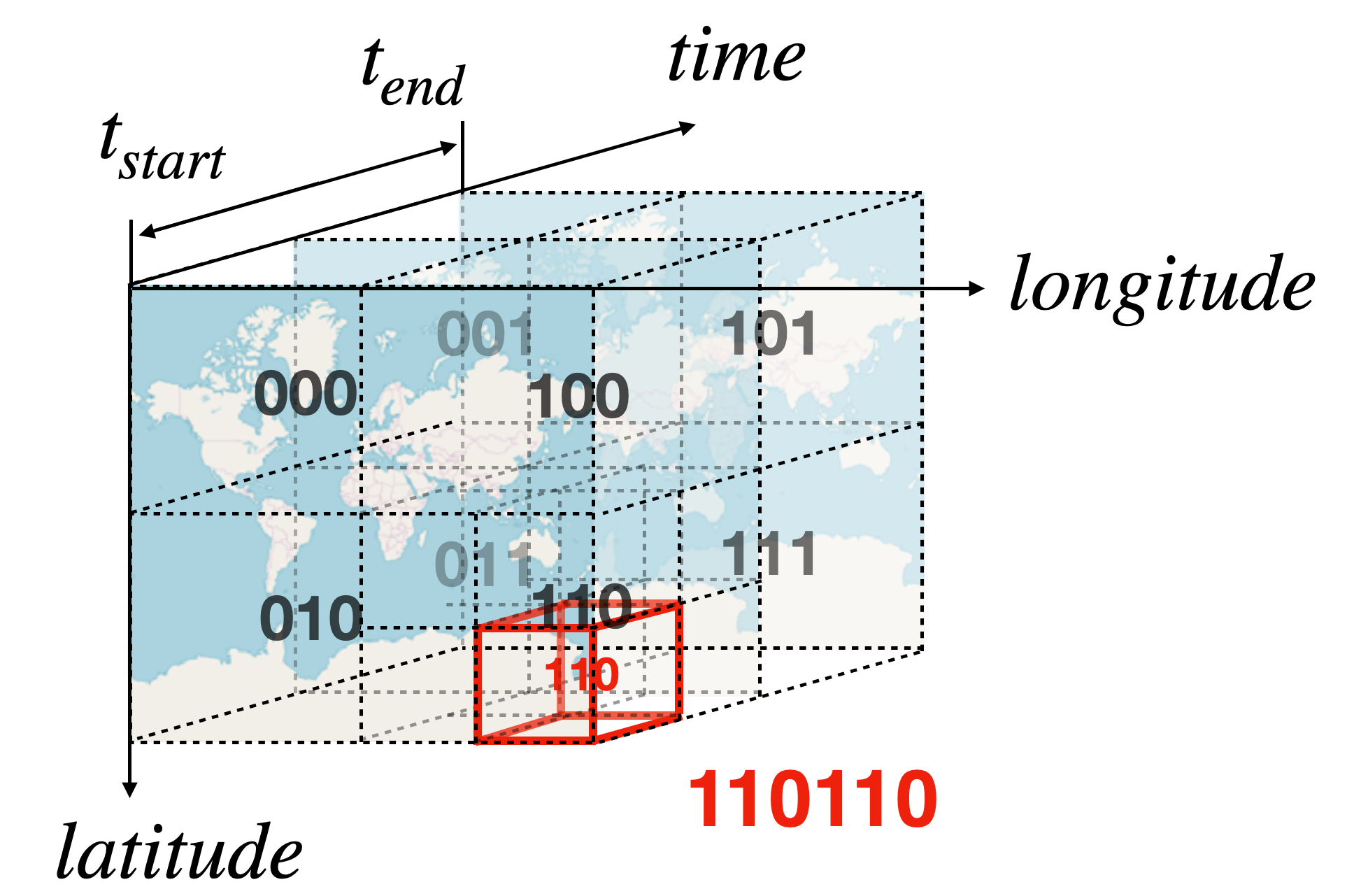}
    \caption{TrajectoryHash interpretation in the time-series map.}
    \label{fig:mixing3d}
\end{figure}

Now, we have three binaries, i.e., this longitude: 1110000000111010, latitude: 0110100100111110 and periodic: 0011100100100.
We mix them into one binary by {\textsc{BitMix}} (line 3).
We consider that there can be some variants, mixing or simply merging without mixing.
A plausible option is to mix one by one from each binary as shown in Figure \ref{fig:mixTH}.
In this mixing, the 3D trajectory data are encoded as in Figure \ref{fig:mixing3d}, where the 3D similarity of the trajectory data is naturally preserved in the binaries in a balanced manner regarding time and location.
By changing the mixing, we can further generalize the encoding.
For example, in the encoding of our previous work \cite{kato2020secure}, the spatial and temporal information were merged sequentially (Figure \ref{fig:seqTH}) to efficiently share and store the prefixes on the nodes of a finite state automaton.

Last, we encode the bits into bytes by {\textsc{ByteEncode}} for ease of transport and processing.
This may involve extra padding in the prefix to match the size of the bytes. 
However, since the padding is common in the prefixes, it is largely ignored in the FST.

Here, we explain that the FST satisfies our three abovementioned requirements.
We can use the FST as key-based data to store and compress the byte sequence data, sharing prefix bytes from the tree structure's roots.
The FST also provides a fast key-based search as a dictionary in proportion to the maximum depth. 
Moreover, the search cost can be $\mathcal{O}(1)$ if the maximum length is small, which is asymptotically equivalent to the Hashmap and may be advantageous because it does not require computing hash functions.
Thus, it basically meets our requirements.
In addition, the FST can also provide operations using select and rank to speed up the internal movement between nodes and can efficiently perform multiple adjacent key searches.
Therefore, we increase the compression efficiency and search performance for the PSI by introducing the FST, and this data structure satisfies our requirements.



\subsubsection{Chunking}
\label{sec:5-2}
We have to consider how to make a chunked FST.
At step 2 of Figure \ref{fig:carousel}, we transform raw data into chunked dictionary representations.
Generally, chunking the FST is not a straightforward task because the compression results depend on how we divides the dataset for each chunk, which is different from the Hashmap.
With the FST, the greater the similarity of the data contained in each chunk, the more it can be compressed.
With the Hashmap, the performance does not depend on these processes because the data size is determined just by the number of stored data.
However, we can solve this problem by simple operation.
Before constructing the FST, we sort the encoded values in byte order and iteratively take the $N_D/N_R$ trajectory data from top to bottom and transform the data into a single FST.
Then, we obtain $N_R$ chunk of FST.
We can stably construct well-compressed FST because a chunk of data has more similarity.



\subsection{Complexity Analysis}
\label{sec:5-4}

\begin{figure}[!t]
  \begin{algorithm}[H]
    \caption{Spatiotemporal PSI}
    \label{alg:pct}
    \begin{algorithmic}[1]
      \Require $q_{i}(i=1,...,n_{c})$, $\Theta=(\Theta_{geo}, \Theta_{time})$, $R \leftarrow {\textsc{mapToChunkedDictionary}}(D, \Theta)$ \algorithmiccomment{\circled{1}, \circled{2}}
      \Ensure $Responses$
      \State ${\textsc{loadToEnclave}}(q_1,...,q_{N_C})$ \algorithmiccomment{\circled{5}~\circled{6}}
      \State $q_1,...,q_{n_c} \leftarrow {\textsc{decrypt}}(q_1,...,q_{n_c})$ \algorithmiccomment{by AES-GCM etc. using shared key through RA}
      \State $Q \leftarrow {\textsc{mapToArray}}(q_1,...,q_{N_C})$ \algorithmiccomment{\circled{6}, $q_i$ has list of encoded value and client ID}
      \State $Results \leftarrow \{\}$
      \For{$r_i \leftarrow R$} \algorithmiccomment{\circled{7} $R$ has $N_{D}$ chunks}
            \State ${\textsc{loadToEnclave}}(r_i)$
            \algorithmiccomment{iteratively load chunked data $r_{i}$}
            \State $r_i \leftarrow {\textsc{decrypt}}(r_i)$
            \For{$query$ in $Q$} \algorithmiccomment{\circled{7} $Q$ is array with $N_{Q}$ length}
                \If{$r_i.{\textsc{contains}}(query.value)$}
                    \State $Results \leftarrow Results \cup query.clientID$
                \EndIf
            \EndFor
      \EndFor
      \State $Responses \leftarrow {\textsc{constructResponses}}(q_1,...,q_{N_C}, Results)$ \algorithmiccomment{\circled{8}}
      \State $Responses \leftarrow {\textsc{encrypt}}(Responses)$ 
      \State $Responses \leftarrow {\textsc{loadFromEnclave}}(Responses)$  \algorithmiccomment{return encrypted data to untrusted area}
    \end{algorithmic}
  \end{algorithm}
\end{figure}

\begin{figure}[!t]
  \begin{algorithm}[H]
    \caption{Spatiotemporal PSI (duration of exposure)}
    \label{alg:pct_doe}
    \begin{algorithmic}[1]
      \Require $q_{i}(i=1,...,n_{c})$, $\Theta=(\Theta_{geo}, \Theta_{time})$, $\Theta_{doe}$, $R \leftarrow {\textsc{mapToChunkedDictionary}}(D, \Theta)$
      \Ensure $Responses$
      \State ${\textsc{loadToEnclave}}(q_1,...,q_{N_C})$ 
      \State $q_1,...,q_{n_c} \leftarrow {\textsc{decrypt}}(q_1,...,q_{n_c})$ 
      \State $Results \leftarrow \{\}$
      \For{$r_i \leftarrow R$}
            \State ${\textsc{loadToEnclave}}(r_i)$
            \State $r_i \leftarrow {\textsc{decrypt}}(r_i)$
            \For{$q_i$ in $(q_1,...,q_{N_C})$}
                \State $durationOfExposure \leftarrow 0$
                \For{$query$ in $q_i$}
                    \If{$r_i.{\textsc{contains}}(query.value)$}
                        \State $durationOfExposure++$ \\
                        \algorithmiccomment{add the interval of sampling rate of the trajectory data collection. (e.g., 1 minute)}
                        \If{$durationOfExposure \ge \Theta_{doe}$}
                            \State $durationOfExposure \leftarrow 0$
                            \State $Results \leftarrow Results \cup query.clientID$
                            \State \textbf{break}
                        \EndIf
                        \State $Results \leftarrow Results \cup query.clientID$
                    \Else
                        \State $durationOfExposure \leftarrow 0$
                    \EndIf
                \EndFor
            \EndFor
      \EndFor
      \State $Responses \leftarrow {\textsc{constructResponses}}(q_1,...,q_{N_C}, Results)$ \algorithmiccomment{\circled{8}}
      \State $Responses \leftarrow {\textsc{encrypt}}(Responses)$ 
      \State $Responses \leftarrow {\textsc{loadFromEnclave}}(Responses)$  
    \end{algorithmic}
  \end{algorithm}
\end{figure}

Here, we discuss the asymptotic computational costs of the PSI and precautions.
We show our stPSI algorithm for trajectory-based PCT in Algorithm \ref{alg:pct}.
Some of the functions are described in \ref{sec:4}.
Dictionary $r_i$ must implement the {\textsc{contains}} method, which returns a Boolean value indicating whether the dictionary includes the target or not.
In the case of the FST, it is asymptotically constant.
The computational costs of trajectory-based PCT are as follows.
Assume that the cost of a single key search for a dictionary is $c$ and that the unique size of $Q$ is $N_{Q}$.
The calculation cost is
\vspace{-3px}
\begin{equation*}
c \times N_{R}\times N_{Q} = \mathcal{O}(N_{R}N_{Q})
\vspace{-2px}
\end{equation*}
Seemingly, $N_{Q}$ and the number of chunks $N_{R}$ is constant, and the PSI is completely scalable for an infected trajectory size.
However, note that the size of $N_{R}$ depends on the memory constraints of SGX.
When processing thousands of queries together, the exact $q_i(i=1,...,N_{C})$ information needs to be kept within the enclave to correctly reconstruct the response, which can be several tens of MB in size; eventually, the size available for chunk $r_i$ is not large.
This means that there is actually a practical lower bound on $N_{R}$.
Last, our routine includes {\textsc{decrypt}} and {\textsc{encrypt}}.
These encryptions are implemented by fast and simple methods, such as 128-bit AES-GCM, and the HW module for encryption is used inside SGX so that the execution time is not dominant.

Algorithm \ref{alg:pct_doe} shows the algorithm for considering the duration of exposure.
The difference is that $\Theta_{doe}$ is required as input, and the number of cases where the PSI is positive are counted continuously for each client (i.e., duration of exposure).
As on line 11, if the server data contain client data, then we add the interval of the sampling rate of the trajectory data collection to $durationOfExposure$.
If the server data contain continuous client data, $durationOfExposure$ will be increased; otherwise, it will be reset to zero.
Finally, the client is considered positive only when the duration of exposure to risky contacts exceeds $\Theta_{doe}$.
Note that this algorithm requires only one scan of the query data; thus, the computational complexity is the same as that of Algorithm \ref{alg:pct}.

\subsection{Accuracy Analysis}
\label{sec:5-5}
\begin{figure}[t]
 \begin{minipage}[t]{0.45\hsize}
  \begin{center}
    \includegraphics[width=\linewidth]{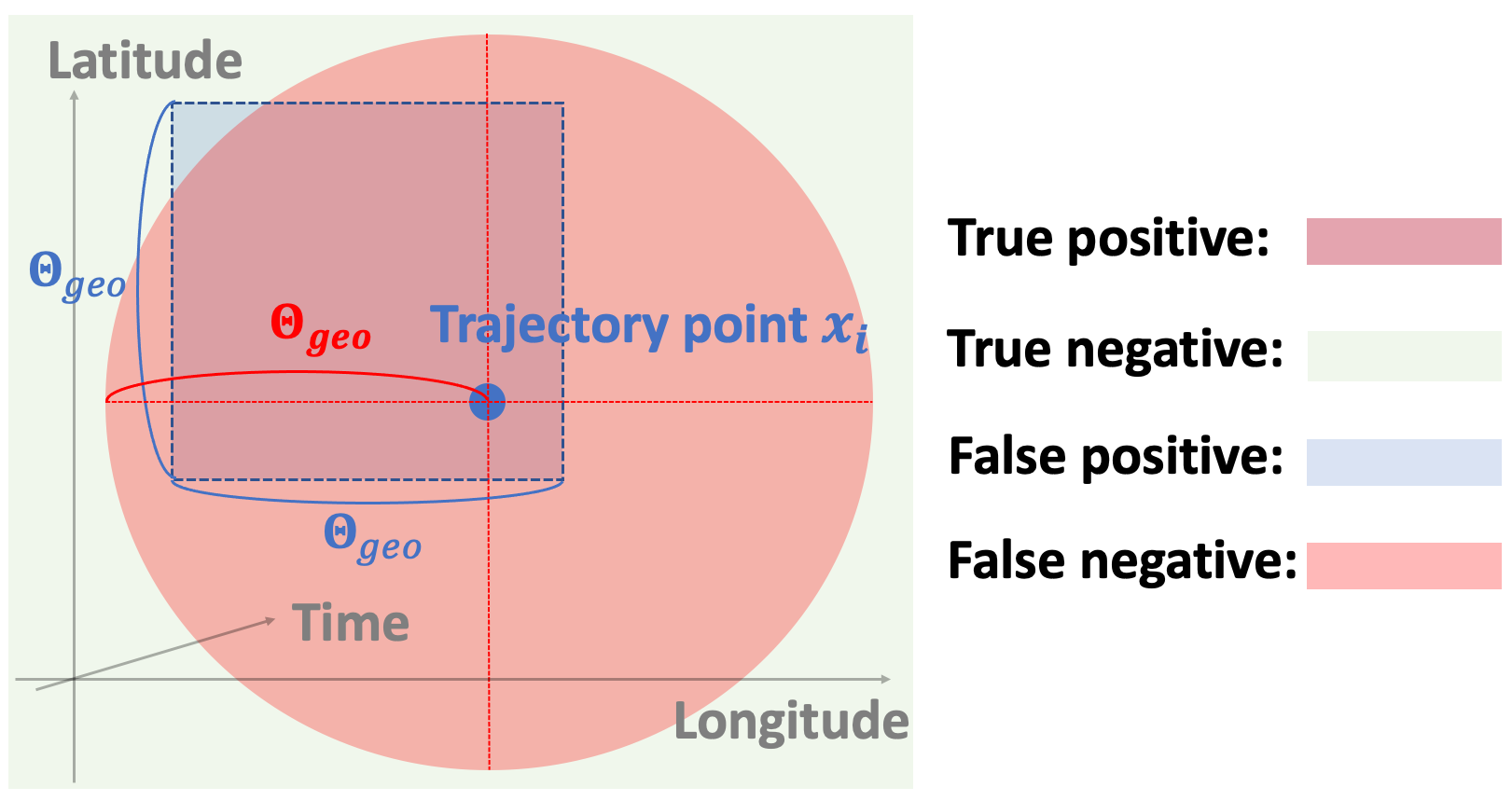}
  \end{center}
    \caption{Possible false outcome in terms of the location.}
    \label{fig:false_geo}
 \end{minipage}
 \hspace{0.1cm}
 \begin{minipage}[t]{0.45\hsize}
  \begin{center}
    \includegraphics[width=\linewidth]{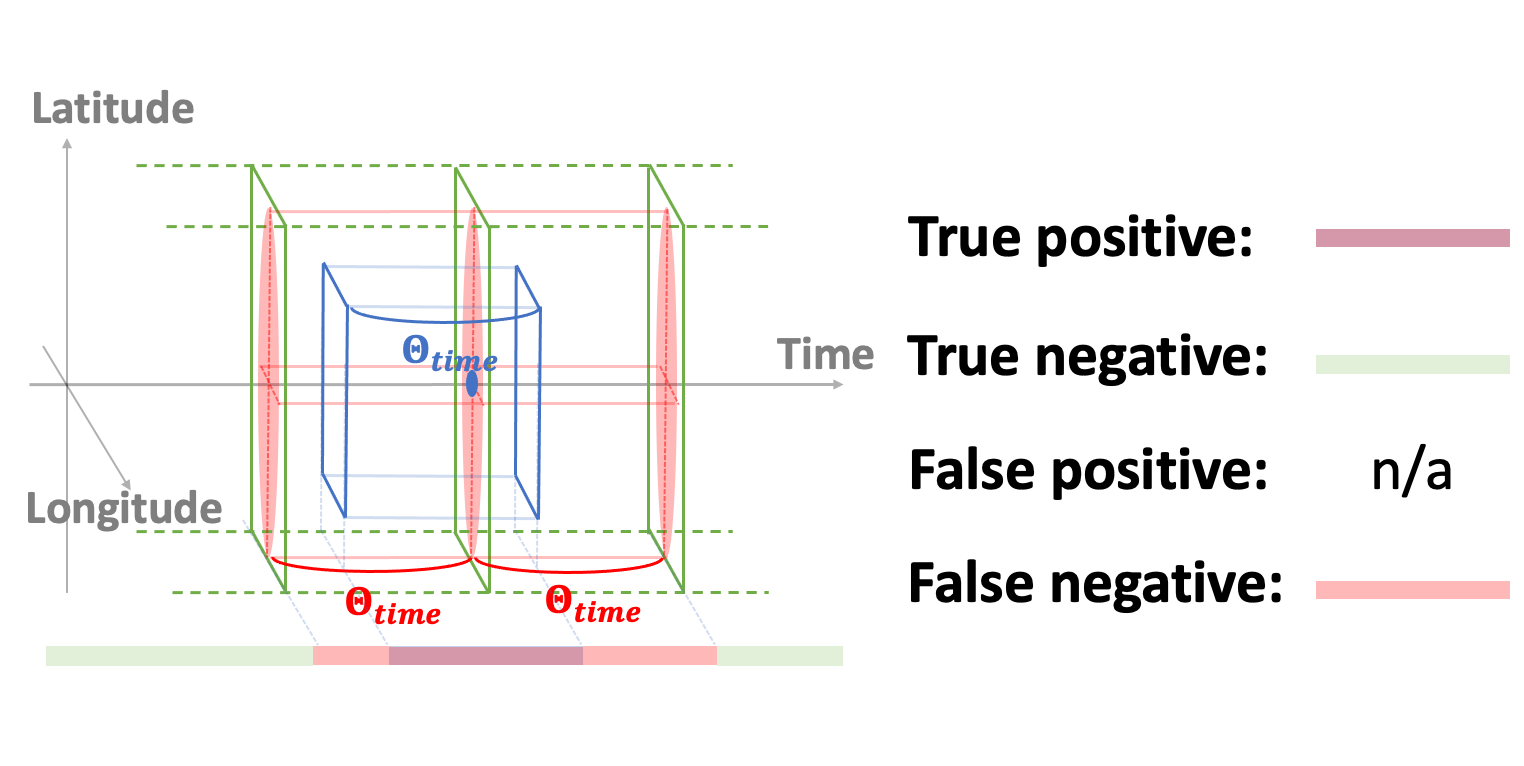}
  \end{center}
    \caption{Possible false outcome in terms of time.}
    \label{fig:false_time}
 \end{minipage}
\end{figure}

Here, we analyze how accurate a contact decision by the stPSI (i.e., Eq.(\ref{eq:3})) is compared to the correct contact decision (i.e., Eq.(\ref{eq:1})).
Since Eq.(\ref{eq:3}) is an approximation of Eq.(\ref{eq:1}), false negatives and false positives may occur.
In Figures \ref{fig:false_geo} and \ref{fig:false_time}, the central blue point represents one trajectory data point $x_i$ included in the query from the client. 
The blue rectangle represents the area of the TrajectoryHash value to which $x_i$ belongs (i.e., the stPSI results are positive if there is infected person data in the same area). 
The red circle represents the area that is judged positive by the exact PCT (i.e., Eq. (\ref{eq:1})).
In particular, Figure \ref{fig:false_geo} shows the projection of 3D space onto the longitude and latitude dimensions, indicating a possible false pattern.
The false positive occurs in the blue area on the upper left of the blue rectangle in the figure, depending on the position of the trajectory point in the blue rectangle.
If there is a false positive, the distance between the false positive data and the trajectory point is between $\Theta_{geo}$ and $\sqrt{2}\Theta_{geo}$.
A false negative can occur over a relatively large area, as indicated by the red circle in the figure.
The possible distance from the trajectory point ranges from 0 to $\Theta_{geo}$.
Figure \ref{fig:false_time} shows the occurrence of false results in the time direction, where the blue point is the trajectory point, the blue area is the area judged as positive by the stPSI, and the red area is the area judged as positive by Eq.(\ref{eq:1}).
In the time-axis direction, since the amount of $\Theta_{time}$ that is judged positive by the stPSI is always contained within $2\Theta_{time}$, it is judged positive by Eq.(\ref{eq:1}), as shown in the figure.
Therefore, a false positive does not occur, but a false negative occurs.
The bounds of the distance between the possible false negatives and the trajectory point are from $0$ to $\Theta_{time}$.
Note that these analyses are for individual data points.
In an actual contact detection query, we have a large amount of infected person trajectory data.
Therefore, these false outcomes may not be very problematic because the query will return 1 if at least one positive sample is found in the entire infected person dataset.
In Section \ref{sec:6-2}, we perform an empirical evaluation of these false outcomes by experimenting with the trajectory data and show that the resulting detection is highly accurate.

\begin{figure}[t]
 \begin{minipage}[t]{0.6\hsize}
  \begin{center}
    \includegraphics[width=\linewidth]{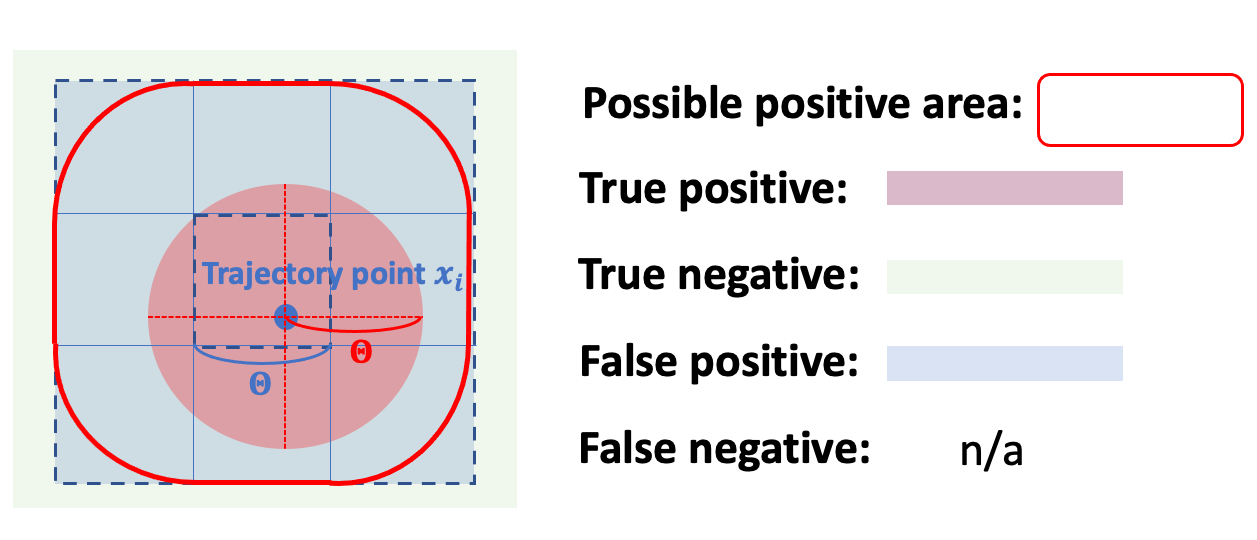}
  \end{center}
    \caption{{\itshape nfp-stPSI} design (2D version for simplicity).}
    \label{fig:false_acc}
 \end{minipage}
 \begin{minipage}[t]{0.35\hsize}
  \begin{center}
    \includegraphics[width=\linewidth]{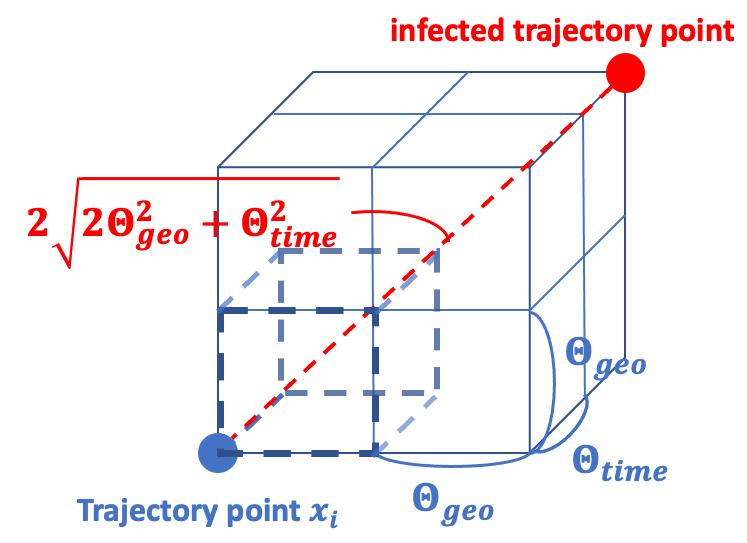}
  \end{center}
    \caption{Maximum distance between trajectory point and false positive.}
    \label{fig:false_max}
 \end{minipage}
\end{figure}

False negatives can occur in the stPSI, which can be problematic.
False negatives in PCT may encourage potentially infected individuals to become active, which may contribute to the spread of infection.
In addition, as shown in the previous analysis, the minimum distance between the trajectory point and the false negative data point obtained by the stPSI is 0.
On the other hand, false positives may be effective in spreading infection.
Pandl et al. \cite{pandl2021detection} shows the importance of appropriate selection of the PDR, suggesting that a wider PDR, i.e., the allowance of false positives, may help prevent the spread of infection.
Therefore, we believe that a false positive is more acceptable than a false negative in PCT.
We devise a variant of the stPSI, {\itshape nfp-stPSI} (non-false negative-stPSI), which has only false positives and no false negatives.
Moreover, there is a guarantee that the distances between the trajectory point and false positive samples are tightly bounded.
nfp-stPSI is a simple extension of the stPSI that intersects all blocks around a block trajectory point belonging to a 3D encoded space (Figure \ref{fig:over_view}).
In other words, the search is performed on 26 surrounding blocks in addition to the original center block to which the trajectory point belongs.
The hash values of these blocks can be efficiently computed from the binary of the trajectory point's TrajectoryHash by some bit operations.
Figure \ref{fig:false_acc} shows the design of the nfp-stPSI in 2D space for simplicity.
nfp-stPSI judges all the surrounding blocks as positive.
Therefore, the exact positive area according to Eq. (\ref{eq:1}), indicated by the red circle in the figure, is always detected as positive.
The area shown by the red bold line in the figure includes all possible exact positive areas (Eq.(\ref{eq:1})) for the trajectory points belonging to the center block, and it is always included in the area of the external blue blocks.
Hence, in nfp-stPSI, false negatives do not occur.
As shown in Figure \ref{fig:false_max}, the maximum distance between the false positives and the trajectory point is reached when the trajectory point belongs to the inner corner of the center block and the infected data are in the outer corner of the external block, with a maximum value of $2 \sqrt{2\Theta_{geo}^2+\Theta_{time}^2}$.
Therefore, the false positive that occurs is guaranteed to be close to the correct answer to some extent.
Thus, although nfp-stPSI increases the overhead of the search, it eliminates false negatives and generates acceptable false positives.
In the next section, we evaluate this method empirically.

\section{Experiments}
\label{sec:6}
We conduct experiments using real trajectory data to demonstrate that the proposed architecture for PCT can achieve high query throughput and the expected properties.

\noindent
{\bfseries Experimental setup.}
We use an HP Z2 SFF G4 Workstation, with a 4-core 3.80 GHz Intel Xeon E-2174G CPU (8 threads, with an 8 MB cache), and 64 GB RAM, which supports the SGX instruction set and has 128 MB processor reserved memory (PRM) in which 96 MB EPC is available for user use.
The host OS is Ubuntu 18.04 LTS, with Linux kernel 5.4.0-72. We use version 1.1.3 of the Rust SGX SDK\footnote{https://github.com/apache/incubator-teaclave-sgx-sdk} \cite{wang2019towards} which supports Intel SGX SDK v2.9.1, and Rust nightly-2020-04-07.
Our experimental implementation is available on Github\footnote{https://github.com/ylab-public/PCT}.

\subsection{Preliminary Experiments}
\label{sec:6-1}

\begin{table}[t]
\centering
\input{tables/psi-ex-balanced}
\label{tab:psi-ex-balanced}
\end{table}

Before the experiments, as described in Section \ref{sec:2-2}, we consider the secure-hardware-based PSI to be much better than the cryptography-based PSI in terms of efficiency.
To verify this, we compare both PSI executions under setting similar to our scenario.
For fairness, we compare single end-to-end PSI query without multiplexing optimization, as described in Section \ref{sec:4}.
Our secure-hardware-based approach implementation is based on Intel SGX and simply uses Hashmap and performs the PSI inside the enclave; the OT-based approach implementation \footnote{https://github.com/osu-crypto/libPSI} follows \cite{rindal2017malicious}.
Table \ref{tab:psi-ex-balanced} describes the execution time comparison between the OT-based \cite{rindal2017malicious} and secure-hardware (Intel SGX) -based PSI under the balanced setting, where we assume that only the RA protocol is performed in advance and that the online phase includes the client-side encryption and decryption time.
We change the set size $10^3$ to $10^6$, and each data point has 128 bit.
As shown in Table \ref{tab:psi-ex-balanced}, Intel SGX can easily overcome the state-of-the-art method in the balanced setting.
In particular, at $10^5$~$10^6$ the difference in execution time becomes significant because of the overhead of oblivious transfer while SGX has scalability in this range of sizes.
Additionally, the secure hardware substantially improves the bandwidth. 
The communication cost of SGX is almost the same as the original size because the data we have to send are just data encrypted by a symmetric key such as AES-128.
Assuming many clients, this condition is essential.
Although the efficiency is better in the two aspects, we should also pay attention to the last line in the table.
Despite using Intel SGX, the execution time is very slow because of the memory constraint of SGX.
When $N = 5.0\times 10^6$ (80 MB), the trusted enclave has to handle approximately 160 MB of data, which exceeds the memory limitation (=96 MB).
As a result, serious overhead occurs.

\begin{table}[t]
\centering
\input{tables/psi-ex-unbalanced}
\label{tab:psi-ex-unbalanced}
\end{table}

Table \ref{tab:psi-ex-unbalanced} shows the results obtained when using Intel SGX in the unbalanced setting.
We also show the results of \cite{chen2017fast} as a reference \footnote{\cite{chen2017fast} does not offer open-source implementation; thus we directly compare with the results they reported.} i.e., the total execution time (online sending and receiver's encryption and decryption) of the best parameters and the maximum multithreading ($ \le 64$).
These numbers are the best of their implementation, but the table shows Intel-SGX-based PSI is significantly fast and efficient in the unbalanced setting, even though the security model is stricter than \cite{chen2017fast} (semi-honest).
The secure-hardware-based PSI is basically not affected by the server-side data size $N$, as shown in Table \ref{tab:psi-feature}.
However, when it is beyond the SGX memory constraint, the execution time becomes slow due to paging overheads, as shown at $N=2^{24}$ (268 MB).
In this case, the client size is very small, and less paging is required, and the impact is smaller than that of the previous result.

In this way, we can achieve a fast PSI by utilizing secure hardware.
We expect cryptography-based methods to gradually improve; however, it is unlikely that they will catch up to the secure-hardware-based method in the near future.
For deployment in a practical situation for the PCT system, it is better to adopt secure hardware.

\subsection{Experiments}
\label{sec:6-2}

\begin{figure}[t]
 \begin{minipage}{0.32\hsize}
  \begin{center}
    \includegraphics[width=\linewidth]{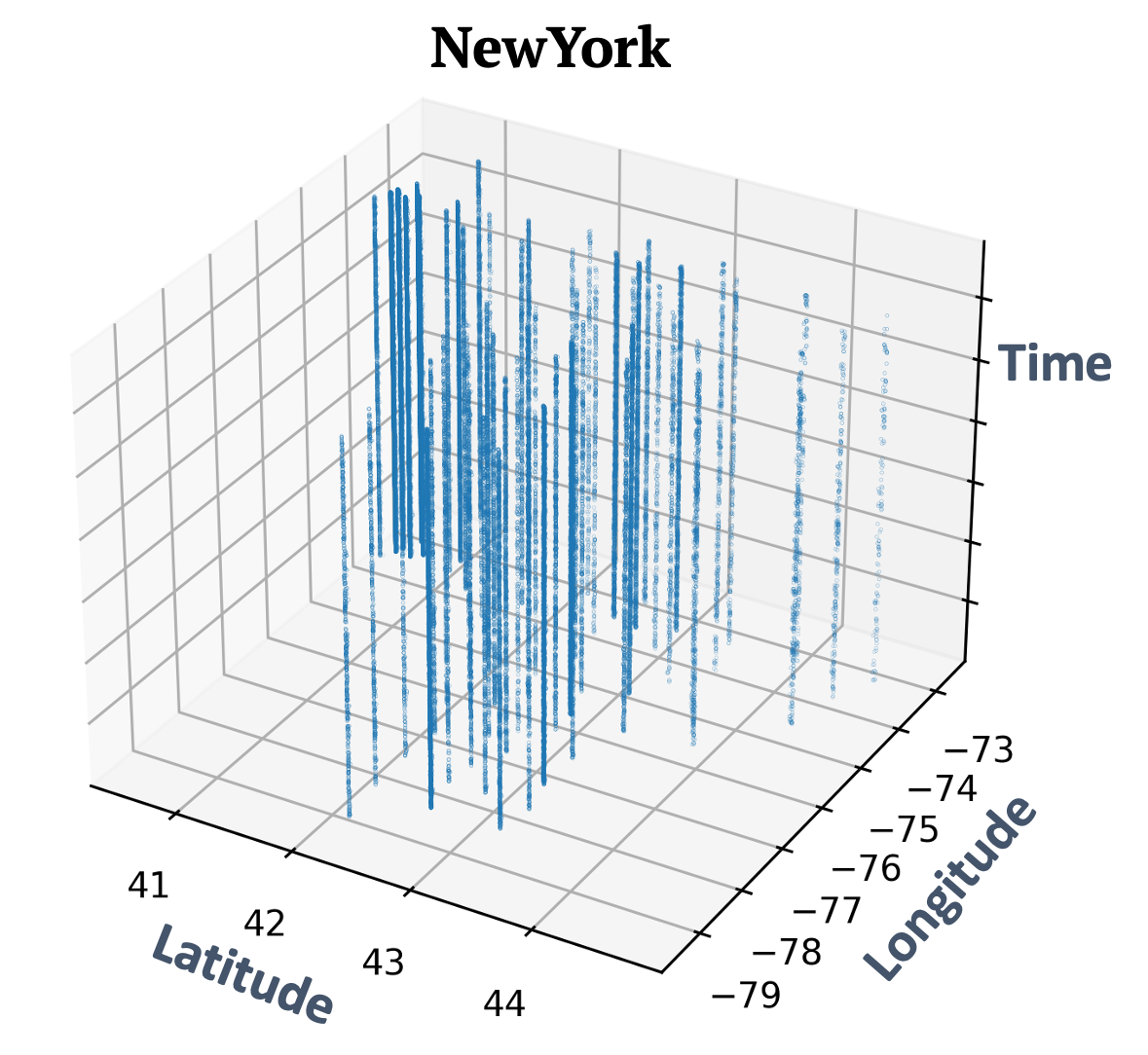}
  \end{center}
 \end{minipage}
 \begin{minipage}{0.32\hsize}
  \begin{center}
    \includegraphics[width=\linewidth]{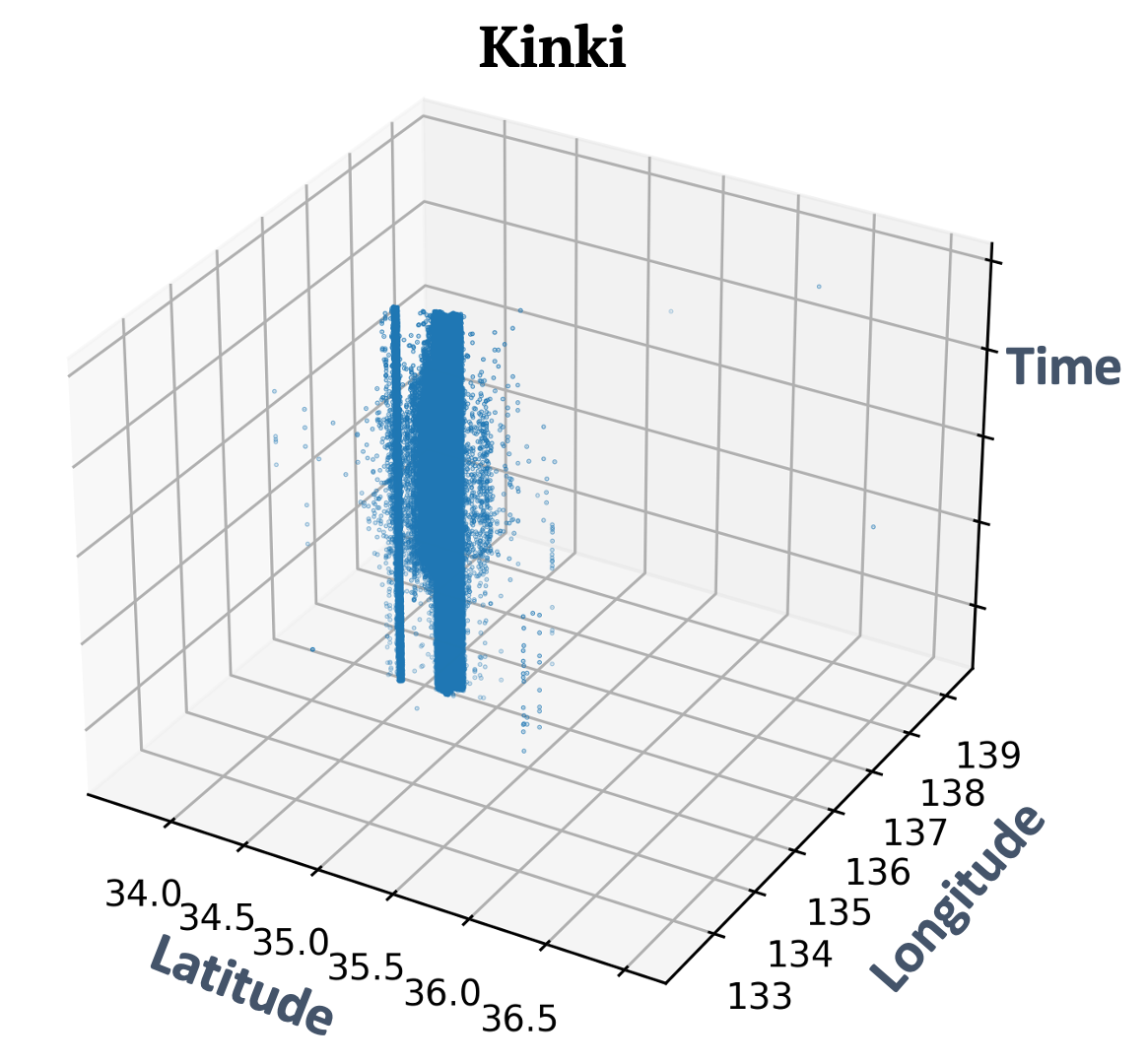}
  \end{center}
 \end{minipage}
 \begin{minipage}{0.32\hsize}
  \begin{center}
    \includegraphics[width=\linewidth]{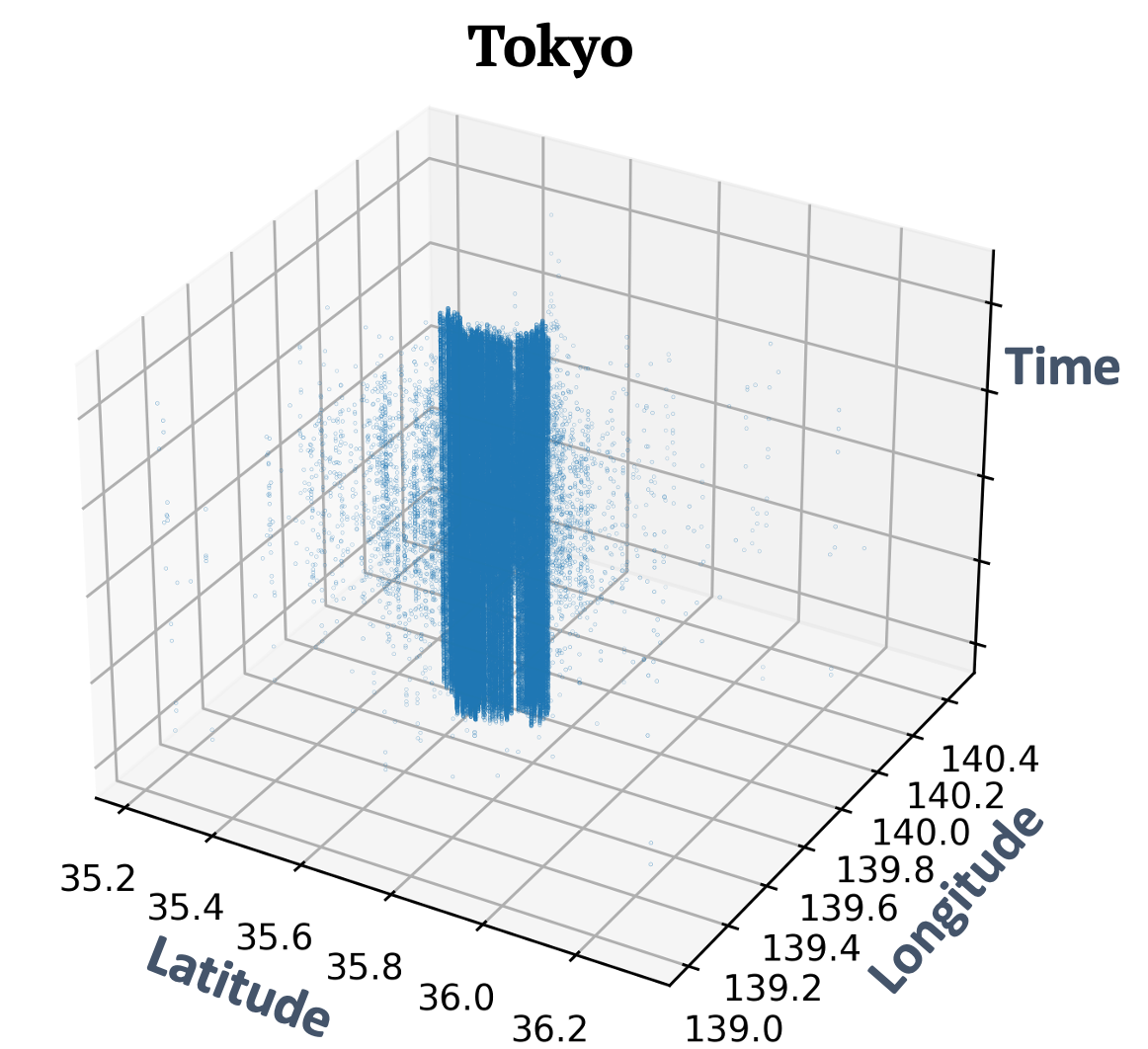}
  \end{center}
 \end{minipage}
 \caption{Trajectory distributions of New York, Kinki and Tokyo. Note that the latitude and longitude scales are different.}
 \label{fig:dist}
\end{figure}

\noindent
{\bfseries Datasets.}
We conduct the experiments with a synthetic dataset and real datasets.
The synthetic dataset is generated by the density EPR model \cite{pappalardo2015returners} implemented in scikit-mobility \footnote{https://github.com/scikit-mobility/scikit-mobility}.
We extend this implementation to describe a more continuous human mobility; it is reproducible by our open-source code \footnote{https://github.com/ylab-public/PCT/tree/master/tools/trajectory}.
The data are individual trajectories for every minute of 14 days in New York City.
The real dataset includes data on people's trajectories in specific regions of Japan available in JoRAS\footnote{http://www.csis.u-tokyo.ac.jp} of The University of Tokyo.
We use the people flow dataset\footnote{Precisely, these people flow datasets are synthetic datasets that are elaborated from real trajectories. More details about the specific process can be found here: \url{http://pflow.csis.u-tokyo.ac.jp/data-provision-service/about-people-flow-data/} and/or \cite{sekimoto2011pflow}.
In our paper, we consider these datasets as real datasets.} for Kinki\footnote{https://joras.csis.u-tokyo.ac.jp/dataset/show/id/3038201000} and Tokyo\footnote{https://joras.csis.u-tokyo.ac.jp/dataset/show/id/3000200800} in Japan to create our experimental dataset.
Note that this dataset has individual trajectories for every minute of only a single day for privacy.
We extract only the time and coordinate information and create our dataset by applying the encoding described in Section \ref{sec:5}.
Figure \ref{fig:dist} shows this 3D trajectory data distribution of New York, Kinki and Tokyo from left to right.
The scatter is 100000 trajectory points randomly sampled.
Since the New York data are generated by EPR model \cite{pappalardo2015returners}, they are distributed in approximately 70 regions.
Compared to Kinki, Tokyo has a much denser distribution.

\subsection{Accuracy Experiments}
To empirically evaluate the contact detection accuracy of the stPSI and nfp-stPSI, we prepare the server-side infected person data and client-side query data for the New York, Kinki and Tokyo datasets.
Table \ref{tab:acc_data} shows the dataset scale.
To prepare these data, regardless of whether they were clients or servers, for the New York dataset, we randomly generated trajectories for each user as abovementioned, and for the Kinki and Tokyo datasets, we randomly sampled users without replacement from original data.
We measure the results of queries by the stPSI and nfp-stPSI and compare them with the correct answers that exactly satisfy Eq. \ref{eq:1} for each client.
The result of each query is obtained as a binary.

\begin{table}[t]
\centering
\input{tables/accuracy_dataset}
\label{tab:acc_data}
\end{table}

Tables \ref{tab:acc_no} and \ref{tab:acc_acc} show the results of the stPSI and nfp-stPSI, respectively.
The tables show the parameters used for the stPSI and the true positive, true negative, false positive, and false negative rates of the contact detection.
Each value represents a percentage and (the number of queries).
The stPSI has a detection accuracy (true positive + true negative) of approximately 90\% for the New York dataset and a higher accuracy, closer to 100\%, for the real Kinki and Tokyo datasets.
However, looking at the New York dataset, the stPSI shows nonnegligible false negatives, as theoretically expected, and the recall is higher, especially as the threshold parameters become more granular.
Compared to the stPSI, nfp-stPSI improves the false negative rate and accuracy for all datasets and causes false positives.
Especially, in the case of New York and $(\theta_{geo}, \theta_{time})=(25, 25)$, nfp-stPSI greatly improves on the true positive and false negative rates, with a relatively small false positive rate.
Moreover, all of these false positives are ``close'' errors.

Overall, the accuracy is high, especially for real data.
Even when the true positive and true negative rates are extremely high in Kinki and Tokyo, respectively, the detection accuracy is high in both cases, which emphasizes that our proposed method is significantly effective.
In the case of real data, we believe that infected people are moving continuously and are more likely to be in contact with multiple points than with a single trajectory data point; hence, the contacts can be captured even with a fixed block area in the stPSI or nfp-stPSI.

We implement the method in \cite{Reichert2020PrivacyPreservingCT} and experiment with its accuracy.
The method performs PCT based on trajectory data, similar to our method, but considers only the location information and not the temporal proximity.
Therefore, we randomly select 100 time points out of 14 days and apply their method on the trajectory data of those time points.
In their method, each data point is rounded to the closest position on the grid, where the distance between the points on the grid is set by $\theta_{geo}$ and $\theta_{time}$ in our experiment.
As shown in Table \ref{tab:acc_obliv}, their method can generate false negatives as well as the stPSI can.

Note that in these accuracy evaluation experiments, the accuracy of collecting trajectory data including GPS tracking is beyond the scope.
The detection accuracy we are discussing is focused only on the error generated by the algorithm, not the error caused by the radio signal troubles, and so on.

\begin{table}[t]
\centering
\input{tables/accuracy_normal}
\label{tab:acc_no}
\end{table}

\begin{table}[t]
\centering
\input{tables/accuracy_accurate}
\label{tab:acc_acc}
\end{table}

\begin{table}[t]
\centering
\input{tables/accuracy_obliv}
\label{tab:acc_obliv}
\end{table}

\subsection{Performance Experiments}

We mention two notes about the performance experiment as a whole.
First, note that in the following experiments, both the client and server send trajectory data ``every minute'' for two weeks.
In practical scenarios, the data interval may be larger, and hence the client query size will be smaller.
Therefore, the number of clients and infected people are not very important in these experiments.
Second, in measuring the execution time, we report the worst case of the stPSI, i.e., while we can avoid checking queries that have already been judged as positive in previous chunks during the execution of the stPSI, since the stPSI depends on the distribution of the data, we do not skip this step and always calculate the intersection for all query data.
This is important in terms of security and can be a countermeasure against side-channel attacks, which guess the query result from the execution time.
Hence, the execution time practically be shorter than the experimental result as usual.

\begin{figure}[t]
  \includegraphics[height=130pt]{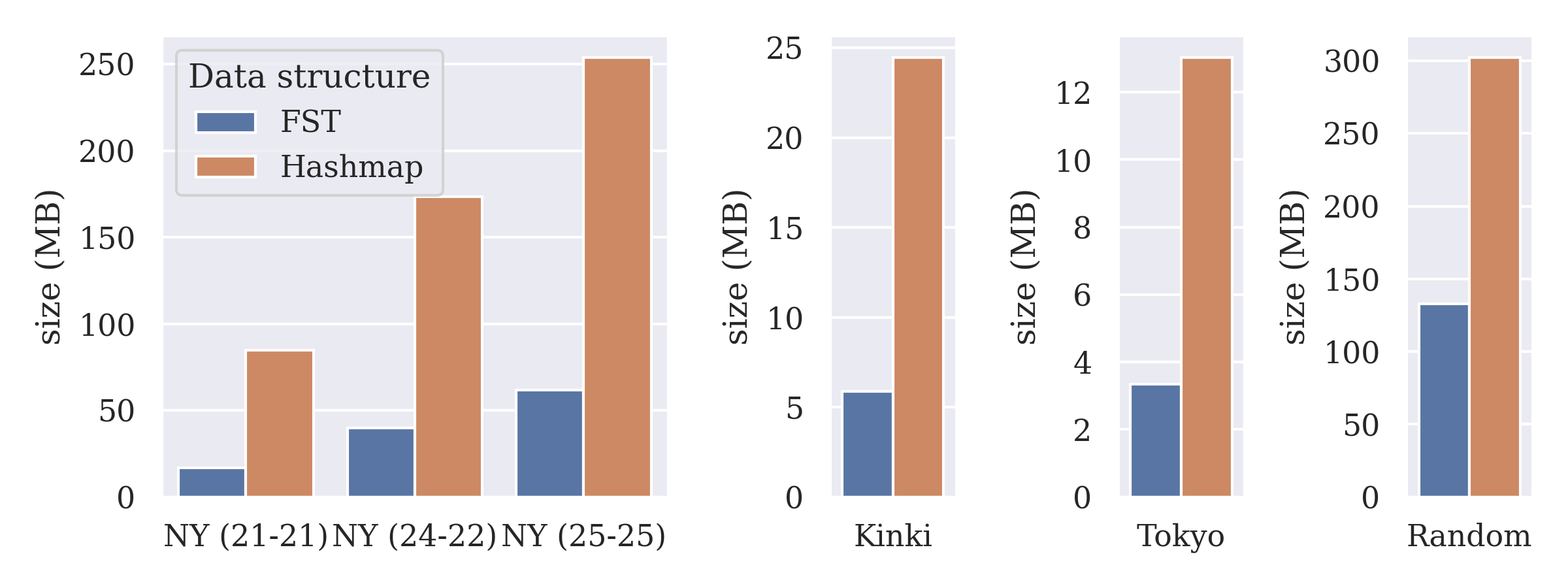}
  \caption{FST can efficiently keep trajectory data encoded by TrajectoryHash for different datasets.}
  \label{fig:compression}
\end{figure}

First, we show the compression results.
In our method, the trajectory data are encoded by TrajectoryHash and stored in the FST.
The baseline method is to use TrajectoryHash and Hashmap (of the state-of-the-art Rust standard library \footnote{https://github.com/rust-lang/hashbrown}).
Figure \ref{fig:compression} shows the sizes of the dictionary data representation for TrajectoryHash-encoded trajectory points stored in the FST and Hashmap.
The dataset used here is the server-side dataset shown in Table \ref{tab:acc_data}, and Random is random data instead of trajectory data.
NY (24-22) corresponds to the New York dataset with TrajectoryHash parameters $(\theta_{geo}, \theta_{time})=(24, 22)$.
The results show that, overall, the FST is able to efficiently compress and keep the data encoded by TrajectoryHash compared to the Hashmap.
For many datasets, the compression is more than approximately 5 times better.
Therefore, we can see that compression can be achieved regardless of the granularity of the parameters or density of the trajectory data.
The fact that the compression efficiency for the actual trajectory data is greater than that for the random data confirms that TrajectoryHash preserves the similarity of the trajectory data and that the FST is able to compress the data well.
The reason why the Hashmap size is different for the same data size is that some of the trajectory data can be grouped into the same TrajectoryHash value (e.g., continuous observations at the same location within a few minutes).
This result supports the fact that the combination of TrajectoryHash and the FST can perform PCT on a large amount of data, even under the tight memory constraints of SGX.

\begin{figure}[t]
 \begin{minipage}[t]{0.45\hsize}
  \begin{center}
    \includegraphics[width=\linewidth]{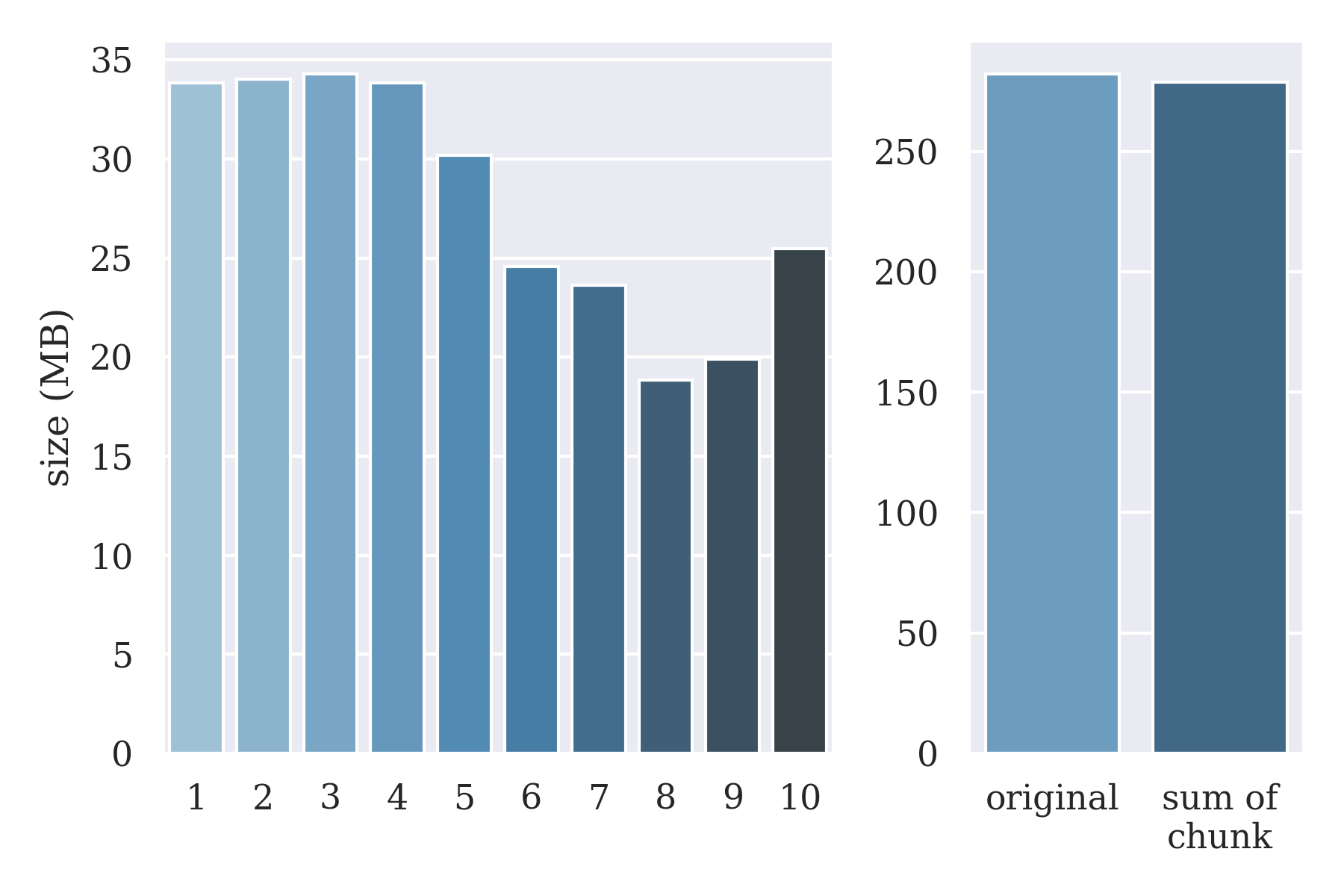}
  \end{center}
    \caption{Sizes of 10 chunks for 282 MB server data, and original data size and sum of these chunks.}
    \label{fig:chunk-detail}
 \end{minipage}
 \hspace{0.1cm}
 \begin{minipage}[t]{0.45\hsize}
  \begin{center}
    \includegraphics[width=\linewidth]{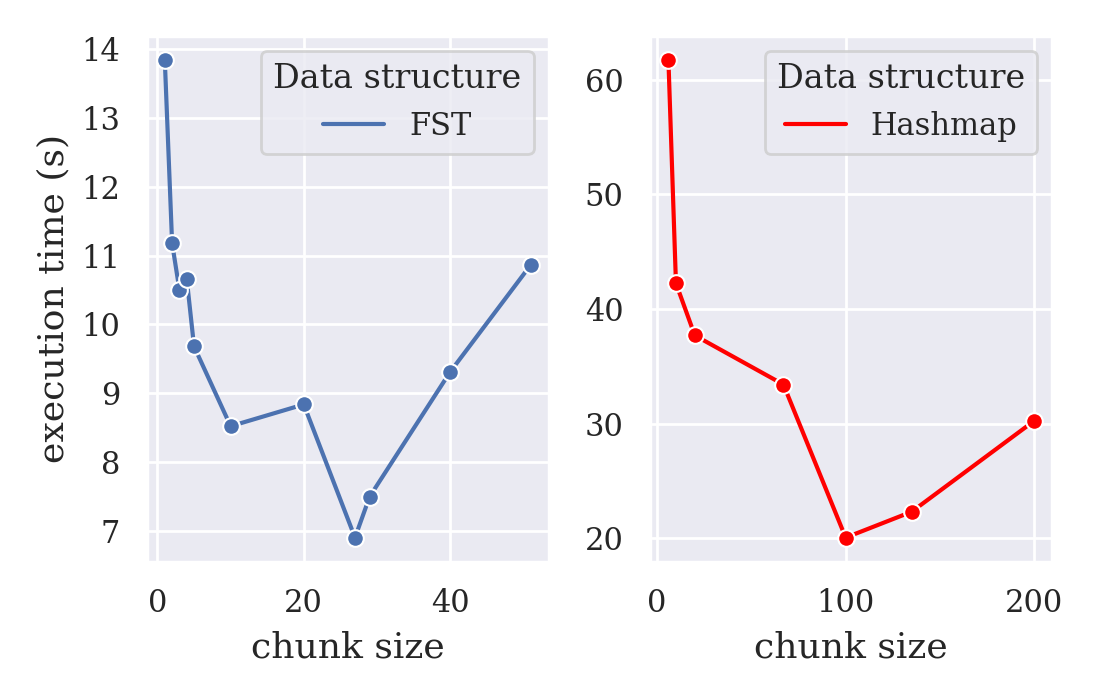}
  \end{center}
    \caption{The execution time varies depending on the number of chunks.}
    \label{fig:chunk}
 \end{minipage}
\end{figure}

To operate our system efficiently under the strict enclave memory constraint, we have to determine the proper chunk size.
We experiment to find the proper chunk size, using the New York dataset and encoding with $(\theta_{geo},\theta_{time})=(25,25)$.
The server-side data size is $20160\times 5000$, and the client-side data is $20160\times 200$.
If the chunk size is 1, i.e., without chunking, the server-side data size is approximately 282 MB, which causes very large overhead in allocating memory in the enclave area because the process tries to allocate virtual memory using special paging beyond the EPC size.
We address chunking the FST by presorting the trajectory data, dividing the data into chunks, and converting each chunk into an FST as described Section \label{sec:5-2}.
Figure \ref{fig:chunk-detail} shows the data for each chunk when the above-introduced server data are divided into 10 chunks.
The sum of the data size is slightly less than the original data size of 282 MB.
We consider each FST to keep data with high similarity by sorting before chunking.

Then, we evaluate the relation between stPSI performance and chunk size.
Figure \ref{fig:chunk} shows the PSI execution time for different chunk sizes when using the FST and Hashmap.
The results of the FST show that the execution time decreases as the chunk size increases, being fastest around 20 to 30, and becomes slower as the chunk size increases.
Compared to the result of the Hashmap, the FST can achieve better performance overall.
When the chunk size is small, the execution time is affected by the memory constraints of SGX, and when the chunk size is large, the execution time is affected by the cost of the repeated ECALL, which is function used to upload data into the enclave and switch the process to a trusted process.
Considering that the client query data size is approximately 32 MB (= 8 bytes $\times$ 20160 $\times$ 200), the results are reasonable since the EPC limitation is 96 MB.
For example, when the chunk size is less than 5, the execution time increases despite the small number of iterations because each chunk is more than 56 MB (=282 MB /5), which incurs a paging cost.
However, when the chunk size is small (e.g., less than 5), there is not such a large overhead even when the EPC memory size is exceeded.
This outcome occurs because the internal implementation of the FST is such that the data to be accessed belong to a contiguous memory area and the number of paging occurrences is small.
In the case of the Hashmap, the overhead is more outstanding.
Although not included in the figure, when the chunk size is 1, the execution time is 532 seconds.
The size of one chunk when divided into 10 chunks is roughly 100 MB, and if divided into 50 chunks, it is approximately 20 MB.
Compared to the FST, since the memory layout of the data of the Hashmap is scattered, the paging of the EPC is likely to occur frequently.

\begin{figure}[t]
  \begin{center}
    \includegraphics[width=\linewidth]{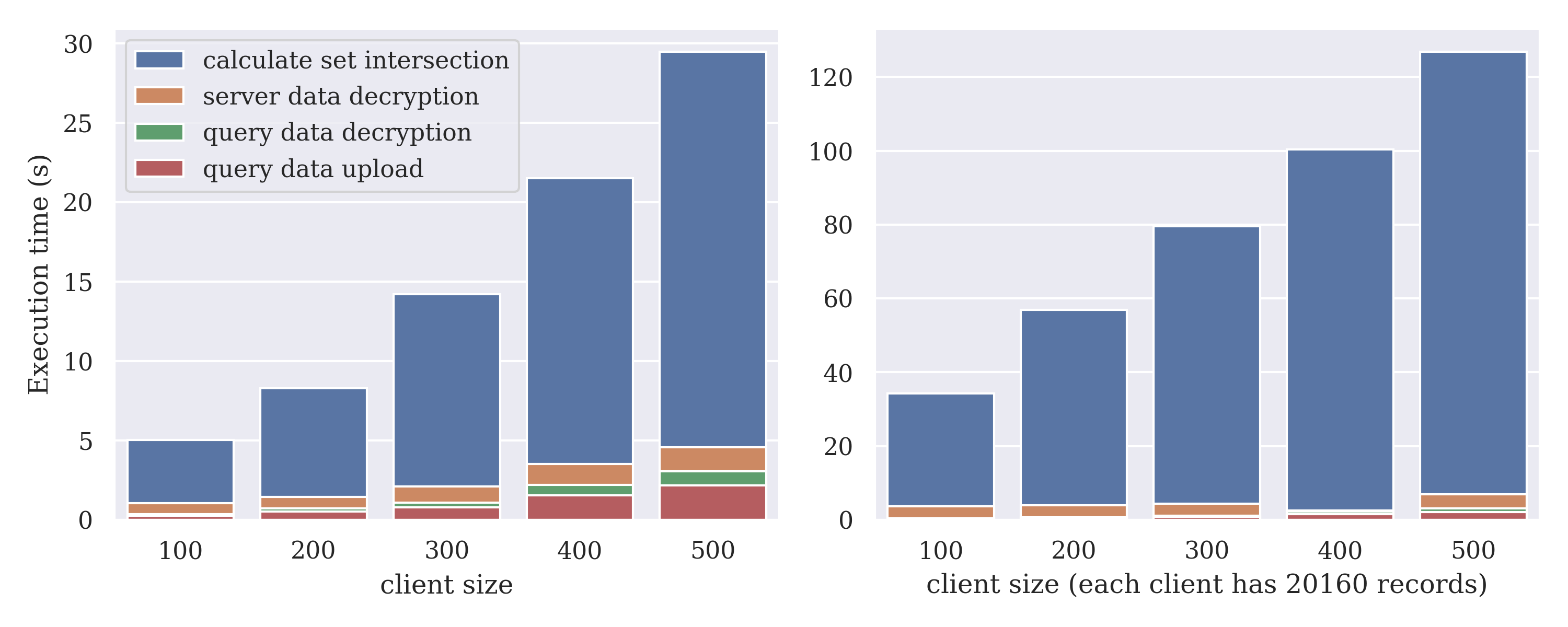}
  \end{center}
 \caption{The stPSI can achieve high performance on a practical data scale (left). nfp-stPSI causes a larger overhead (right).}
    \label{fig:total-executiontime}
\end{figure}

\begin{figure}[t]
 \begin{minipage}[t]{0.36\hsize}
  \begin{center}
    \includegraphics[width=\linewidth]{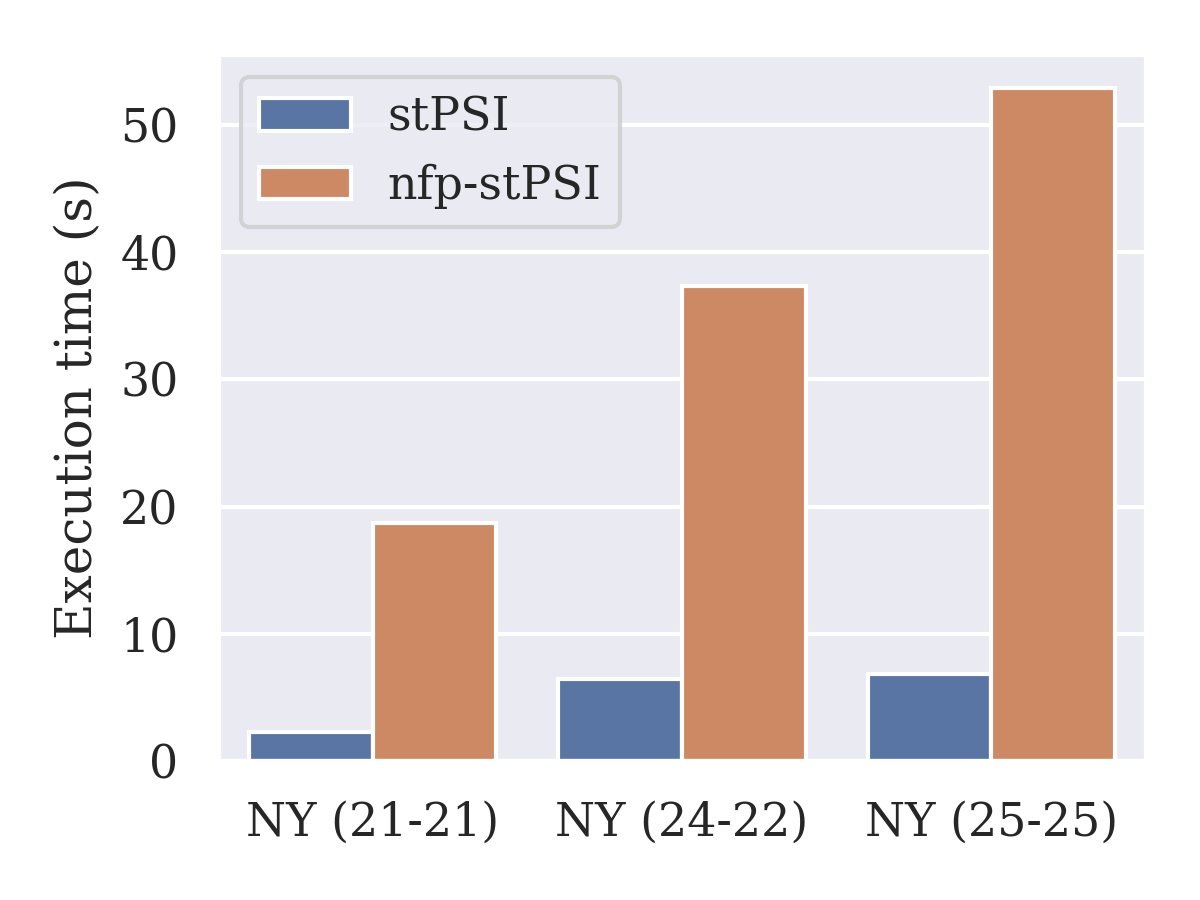}
  \end{center}
   \caption{Finer granularity of parameters directly degrades the performance.}
    \label{fig:param-perfor}
 \end{minipage}
 \hspace{0.1cm}
 \begin{minipage}[t]{0.60\hsize}
  \begin{center}
    \includegraphics[width=\linewidth]{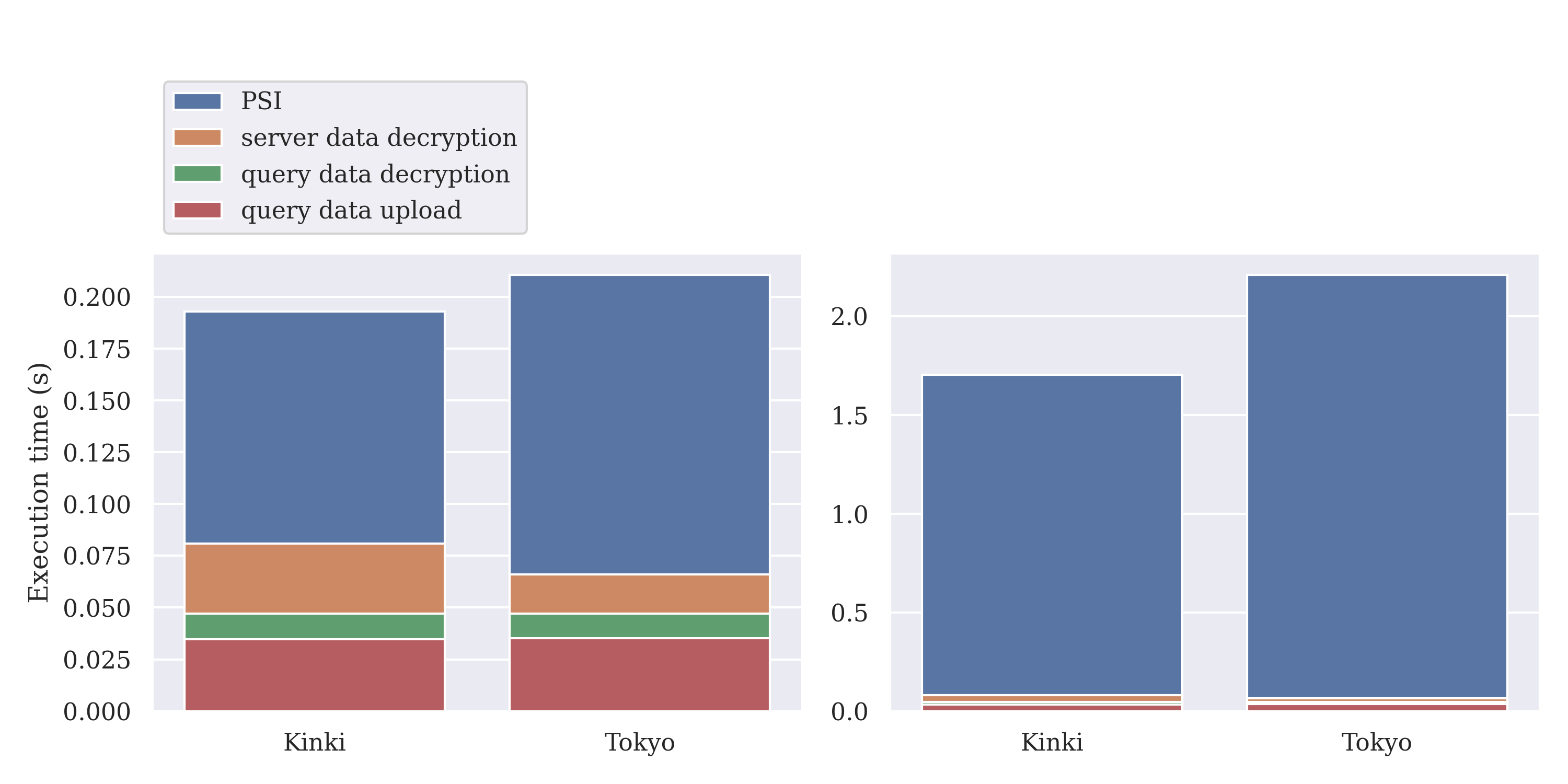}
  \end{center}
   \caption{Results of the stPSI (left) and nfp-stPSI (right) on real data.}
    \label{fig:real-data-perfor}
 \end{minipage}
\end{figure}

Next, we evaluate the overall performance.
Figure \ref{fig:total-executiontime} shows the overall execution time of the PCT when the server data are fixed at 5000$\times$20160 and the client data increase from 100$\times$20160 to 500$\times$20160.
The figure on the left shows the execution time of the stPSI, and the figure on the right shows the execution time of nfp-stPSI.
The execution times of the four phases of the stPSI, i.e., computation of intersection, decryption of the server data, decryption of the query data, and uploading of the query data to the enclave are shown in different colors.
In the stPSI, the number of chunks is set to 20, and in nfp-stPSI, the number of chunks is set to 1.
These values are selected because the number of searches to the FST increases enormously in nfp-stPSI; thus it is more effective to reduce the number of chunks to reduce the number of iterations.
The execution time of the stPSI and nfp-stPSI increases almost linearly with the client data size.
As the client size increases, the time required for uploading and decoding the query data becomes nonnegligible, but the computational cost of the intersection is completely dominant.
Compared with the stPSI, nfp-stPSI causes a large overhead, approximately 4 to 6 times larger.
This result indicates that there is room for further improvement in the implementation of nfp-stPSI.
Overall, however, our proposed algorithms are able to return results with practical execution times for data with a scale close to that of real data.

We compare the results in terms of performance with \cite{Reichert2020PrivacyPreservingCT}, which also achieves PCT by secret computation, similar to our method.
Their method performs the secret computation by ORAM \cite{doerner2017scaling}.
We perform PCT by using the same implementation of binary search with Floram\footnote{https://gitlab.com/neucrypt/floram} as in their work.
We assume that the server data consist of 1000 trajectories and the client data consisted of 100 trajectories\footnote{Note that this is different from the experimental scale of our proposed method because the computation of the competitor is too time consuming.}.
In their algorithm, the precomputation increased the size of the data by a factor of 9, resulting in the need to compute the intersection of 9000 points of server data and 900 points of client data.
Even though this is a setting that is approximately six orders of magnitude smaller than the scale of our experiment, the computation takes 123.3 seconds on average.
This result indicates that the overhead of ORAM is still unacceptable at present and that more emphasis should be placed on the speedup of the secret computation that the TEE brings.

Figure \ref{fig:param-perfor} shows a comparison of the execution time for several granularity parameters.
The used client size is 200$\times$20160, and the server size is 5000$\times$20160.
Because the parameters are set to finer granularity, the execution time essentially increases.
This phenomenon occurs due to the increase in the size of the hash value and the decrease in the compression ratio by the FST due to the decrease in the similarity and coincidence rate of the trajectory data.
The size of the hash value for $\theta_{geo}, \theta_{time}=(21,21)$ is 7 bytes, but for (24,22) and (25,25), it is 8 bytes.

The results of the stPSI and nfp-stPSI execution times for the real Kinki and Tokyo datasets are shown in Figure \ref{fig:real-data-perfor}.
The execution times shown are the same as in Figure \ref{fig:total-executiontime}.
The used client data size is 1440$\times$200, and the server data size is 1440$\times$14000.
In the stPSI, the percentage of the execution time for decryption and query data upload is relatively high, while in nfp-stPSI, the percentage of intersection computation is high.
Overall, the execution time is kept small, which shows that our proposed algorithm works fast even for real data.

\subsection{Requirements Analysis}
We show how our system meets the requirements we introduced in Section \ref{sec:3-2}.

\noindent
{\bfseries Efficiency.}
There are four elements that make the system efficient: Intel SGX, chunking, data representation, and query multiplexing.
First, SGX bring us an efficient PSI.
SGX allows software to perform secret computations transparently and eliminates the need for complicated and time-consuming cryptographic techniques to perform the PSI.
This fact can be confirmed our preliminary experiments and is the main basis of the efficiency of our system.
The computational overhead is small, and also bandwidth is dramatically improved compared to existing methods.
Second, chunking $R$ into $r_i$ $(i=1,...,N_D)$ avoids the serious paging overhead caused by the severe memory constraints of SGX even when the data of infected patients become too large to fit into the enclave.
Third, the memory-efficient dictionary representation (Section \ref{sec:5}) reduces the number of chunks, resulting in reduced PSI execution and overheads for upload to the enclave.
This is also an important element of our system.
Fourth, steps 5 and 6 (Section \ref{sec:4}) show query multiplexing and improve the throughput of the query processing.
Reading the chunked data $r_i$, as in Step 7, is costly due to the $N_D$ iteration, and doing this for every query yields large overheads.
As a result, our system achieves high query throughput and scalability.

\noindent
{\bfseries Security.}
Our protocol follows the RA and secure computation provided by Intel SGX.
Previous studies \cite{bahmani2017secure, subramanyan2017formal} show the protocol security.
Informally, any state cannot be observed from outside the TEE, and even if any inputs are known, any tampering with the state that can be performed by the malicious server will not divulge any information about the client trajectories.
Hence, it is guaranteed that all information an attacker can observe is only outside the TEE.
However, in our system, all information observed outside the TEE must be encrypted.
Therefore, cryptographically strong security for the client's privacy from any external attacker is ensured when using proper encryption.
More formal definitions require elaborate modeling of the attacker and private information, but our setting is common, and we defer to earlier work \cite{bahmani2017secure, subramanyan2017formal}.
Note that some side-channel attacks, such as Spectre \cite{chen2019sgxpectre} and other cache attacks, are beyond the scope of their work and our work.
To protect against these attacks, we have to consider {\itshape data-obliviousness} \cite{mishra2018oblix, ahmad2018obliviate} to make the side channels uniform.

Considering the risk analysis discussed in Section \ref{sec:3-2}, some simple extensions are needed.
DoS attacks: To prevent DoS attacks against SGX, it is necessary to authenticate the client before SGX processes it.
Since the data on the server side change only once a day, each client should be allowed to make only one request per day.
This can be done with a simple extension, since the information of the user requesting the query does not need to be anonymized.
Query-abusing attacks: This can also be done by authenticating the client to minimize the damage of attacks. The client can obtain only the 0 or 1 information.
However, even if the request is limited to once a day, there is a possibility that some information may be leaked as a result of using multiple compromised clients. 
To prevent this, other protection mechanisms are necessary; for example, assuming the possibility of linking the obtained information with other information, preserving differential privacy should be considered.
Side-channel attacks: Since the data are encrypted by a key shared between SGX and the client, it cannot be leaked from the network. On the other hand, attacks through the side channels inside the server may be possible due to cache attacks, etc.
Defending against these attacks is the topic of future work.
False answering: False answering is prevented because the enclave's internal blogs are verified by RA.
Fake data injection: In this scenario, we assume that the health agency is a trusted institution, and when a third party injects data, it is necessary to modify the data in the memory encrypted by SGX. Thus, it is cryptographically protected.
Replay: A replay attack can occur even when a secure channel is in place, but it can be completely prevented by authenticating the user, restricting the number of requests per day, and updating the key every day.

\noindent
{\bfseries Flexibility.}
We achieve flexibility by parameterizing the encoding of the data using a granularity parameter $\Theta$.
The three parameters we introduce are geographic granularity, temporal granularity, and duration of exposure to the virus.
The parameters are shared between the server and clients in advance.
We have to update all the data when we want to change the rules of risky contact.
In other words, our method cannot change the flexibility parameters more frequently than batch updates.
Essentially, we believe that these parameters are determined based on epidemiological arguments and that their change frequency is not more than the frequency of batch updates of the data (i.e., once a day).

\noindent
{\bfseries Accuracy.}
Our proposed method may generate nonnegligible false negatives and false positives.
However, we have shown that it is possible to extend the method to generate only false positives with a reasonable upper bound on the error.
Note that we have focused only on the errors caused by the algorithm.
In the real world, in addition to this, there are errors that occur during data collection and errors that occur due to the physical characteristics of buildings and vehicles.
To perform contact detection based on this information, more metadata and other information may be required.
There is a trade-off between the cost of information collection, privacy, and accuracy.
A more detailed discussion of the trade-offs involved from epidemiological perspectives is beyond the scope.

\section{Related Works}
\label{sec:7}

There are DP3T \cite{troncoso2020decentralized} and similar schemes \cite{becker2019tracking, gvili2020security, rivest2020pact, trieu2020epione}, which are BLE-based decentralized architectures that use a device's wireless signals.
They are the most popular implementation methods to date.
The major difference from our proposed system is that while they handle only contact information through random ID tracking, we directly handle trajectory data in a private manner to detect indirect contacts.
Desire \cite{castelluccia2020desire}, known as hybrid architecture, has basically the similar characteristics as decentralized architectures
In Desire users send random ID-based contact information to the server, but the ID cannot be used to identify an individual.
Therefore, there is no way to discover an indirect contact based on trajectory data.

Hamagen \cite{pinkas2020hashomer} and AAROGYA SETU \footnote{https://www.aarogyasetu.gov.in/} are similar applications developed for COVID-19 but use trajectory data.
Hamagen uses trajectory data to identify the contact location, but the logic of contact determination is different from that of our proposed method because Hamagen is a BLE-based method similar to DP3T.
In other words, Hamagen does not attempt to use trajectory data to detect contacts but rather to publish the locations where contacts with infected people have occurred.
AAROGYA SETU collects raw trajectory data, unlike our system; thus, there are serious privacy concerns \cite{ayyuga}.
There are other studies \cite{culler2020covista, luo2020deepeye} that propose the use of richer data such as users' detailed trajectory data.
In contrast to these studies, our method differs in that we use these rich trajectory data to find instances of indirect contact.

We summarize the comparison between the BLE-based PCT technique and our method (centralized architecture) as follows.
First, from the security and privacy points of view, the characteristics are very different, as described in Section \ref{sec:3-2} and \cite{ahmed2020survey}.
In centralized architectures, it is important to discuss the trust model.
We have to consider how to protect a client's privacy when the server is untrusted.
In the BLE-based approach, there is no such security concern since personal data are not exposed to any party.
On the other hand, BLE-based methods have security risks that do not arise in a centralized architecture, such as continuous tracking of random IDs.
In most cases, to ensure the security of a centralized architecture, either a very costly method such as secret computation or a strong assumption such as trusting the server is required.
This is one of the main reasons why a decentralized architecture is preferred.
Second, in terms of efficiency, as opposed to the BLE-based method, where the contact decision is computed on each device, our method is computed on the server, which can be computationally intensive.
The advantage of the BLE-based method is that the computational cost is lower than that of our method, which compares the data of many infected people, because the target of the computation is limited to only those who have direct contact.
On the other hand, it is not possible to detect, e.g., indirect contact, via the BLE-based method.
In terms of communication cost, the BLE-based method basically requires broadcasting of the infected person data.
While the size of these data is very small, it is necessary to notify all users.
In a centralized architecture such as ours, users need to send information about their trajectory data only when they query the server, but the data size is much larger.
Finally, regarding storage, while the BLE-based method stores only the contact's ID, our method requires the user to store the trajectory data on his own device.
Third, we compare the ability to detect contacts.
We mentioned that our method is able to detect a wider class of contacts (i.e., indirect contacts) by using trajectory data.
However, there are cases where the BLE-based method is more advantageous in the contact detection capability.
Since the BLE-based method relies on wireless signals, contact decisions are performed according to the constraints of the physical environment (e.g., walls, vehicles, etc.).
Therefore, there is a possibility that direct contact can be detected with an accuracy that cannot be captured only by analyzing trajectory data.
This is a noteworthy point compared to our method.

Reichert et al. \cite{Reichert2020PrivacyPreservingCT} proposes a setting similar to that of our study.
The similarity is that they used PCT with secret computation techniques and centralized trajectory data.
Our work differs from theirs in the contact detection algorithm and the secret computation technique used.
The basis of their method is a PSI using ORAM, which has some performance problems.
To compare, we extensively experiment on and evaluate the performance and accuracy of the two methods in this work.

We need to refer to the open-source project SafeTrace (\url{https://github.com/enigmampc/SafeTrace}) \cite{enigma}.
SafeTrace proposes a TEE-based method similar to our proposed method.
SafeTrace focuses mainly on collecting and managing the trajectory data secretly.
Currently, the development is not updated, and the part of contact detection (PCT) and notification service are not yet specified.
Moreover, their work does not include any efficient PCT algorithm.
We develop an algorithm for finding indirect contacts; thus, their work is orthogonal to ours.
One clear difference between our proposed method and theirs is that their system collects the complete trajectory data, while our system uses information that is optimally encoded for PCT and can be applied in the stPSI efficiently.
Hence, our proposed method is able to achieve optimized performance because it is focused on PCT.


A similar type of query to the proximity query is the reachability query \cite{shirani2012efficient, 10.1145/3139958.3139982}.
, where the query is to determine whether or not it is possible to reach a target vertex from a source vertex in a given digraph.
This method has the potential to perform contact tracing by using a graph related to meeting information \cite{10.1145/3139958.3139982}.
However, it is not suitable for answering contact judgment queries for unknown users because it can answer queries only for data that can be precomputed and are already in the database, and it cannot answer unknown data efficiently.
\cite{10.1145/3139958.3139982} has a different feature from our proposed method because it can track contacts on a human basis, not on a location basis.

Regarding trajectory data compression, a well-studied technique is that of Douglas-Peucker, i.e., route-wise compression \cite{song2014press, chen2019trajcompressor, zhao2018rest}.
The basic idea is reproduction by estimating the route from a minimum number of points \cite{song2014press}.
The compression is performed by approximating the route information rather than the position information.
Our compression is based on the similarity of trajectory points.
Therefore, it is orthogonal to these compression methods.
However, this route-wise compression may not work well in contact tracing because it is not possible to tell whether or not points are really in contact just by intersecting.
To detect contact, it is necessary to include some time information in the route information.

\section{Conclusions}
\label{sec:8}
In this paper, we proposed a trajectory-based PCT system using trusted hardware to control the spread of infectious diseases.
We identified the problems of existing PCT systems, clarified the requirements for trajectory-based PCT, and presented a TEE-based architecture to achieve secure, efficient, flexible, and accurate contact tracing.
Our experimental results obtained on real data suggested that our proposed system can work on a realistic scale.
We hope that this study will motivate different communities and help in the development of solutions to combat COVID-19 as soon as possible.

\begin{acks}
This work is partially supported by JSPS KAKENHI Grant No. 17H06099, 18H04093, 19K20269, JST/NSF Joint Research SICORP 20-201031504, and KDDI Foundation.
Additionally, this is joint research with CSIS, the University of Tokyo (No. 974) and used the following data ID: 3000200800, 3038201000.
\end{acks}

\bibliographystyle{ACM-Reference-Format}
\bibliography{bib-base}

\end{document}

%% file: tables/comparison.tex

\begin{tabular}{lccccc}
    \toprule
     & Functionality & Efficiency & Security & Flexibility & Accuracy \\
    \midrule
    \midrule
    DP3T \cite{troncoso2020decentralized} & - & 	\checkmark &  \checkmark & - & \checkmark \\
    HE-based PSI \cite{rindal2017malicious, chen2017fast} & - & - &  \checkmark & - & \checkmark \\
    TEE-based PSI \cite{tamrakar2017circle}  & - & \checkmark &  \checkmark & - & - \\
    MPC-based PCT \cite{Reichert2020PrivacyPreservingCT}  & \checkmark & - &  \checkmark & - & \checkmark \\
    Our work & \checkmark & \checkmark &  \checkmark & \checkmark & \checkmark \\
    \bottomrule
\end{tabular}

%% file: tables/psi-feature.tex
\begin{tabular}{l c c c c} 
\hline \hline 
& {\bf bandwidth} & {\bf computation} & {\bf requirements} & \begin{tabular}{c}
{\bf server} \\ {\bf security}
\end{tabular} \\
\hline
TEE & $\mathcal{O}(n)$ & $\mathcal{O}(n)$ & 
\begin{tabular}{c}
attestation \\ hardware
\end{tabular}
& malicious\\ 
\hline
OT \cite{rindal2017malicious} & $\mathcal{O}(N)$ & $\mathcal{O}(N\log N)$ & nothing &  malicious \\ 
\hline
HE \cite{chen2017fast} & $\mathcal{O}(n\log N)$ & $\mathcal{O}(N)$ & nothing  &semi-honest \\ 
\hline \hline
\end{tabular}
\vspace{3pt}
\caption{PSI comparison: cryptography-based(OT, HE) v.s. TEE-based: TEE needs to special hardware, but it has advantages in efficiency and security.}

%% file: tables/parameters.tex
\begin{tabular}{cl}
    \toprule
    Symbols & Explanation\\
    \midrule
    $N_{D}$ & number of raw trajectory data of infected people \\
    $D$ & raw trajectory data of infected people \\
    $N_{C}$ & number of clients \\
    $c_{i} \in \mathbf{C}$ & a client $i(\in \{1,...,N_{C}\})$ and all clients set \\
    $N_{R}$ & number of chunks of central data \\
    $R$ & mapped $D$, array of efficient chunks $(=(r_1,...,r_{N_R}))$\\
    $r_{i}$ &  $i$-th chunked data of $R$, efficient representation (e.g., FSA) \\
    $q_{i}$ & client $i$'s query data (raw trajectory data) \\
    $Q$ & merged and mapped $N_{C}$ query data (e.g., unique array)\\
    $N_Q$ & unique size of $Q$ \\
    $\theta$ & parameter of PCT, $\theta=(\theta_{time}, \theta_{geo})$ \\
    $(t_{i},l_{i})$ & $i$-th row of trajectory data, time $t_i$ and location $l_{i}=(l_{ilat}, l_{ilon})$ \\
    \bottomrule
\end{tabular}

%% file: tables/param_theta_geo.tex
\begin{tabular}{cl}
    \toprule
    parameter $\theta_{geo}$ & geo distance $\Theta_{geo}$ \\
    \midrule
    26 & 0.6 m \\
    25 & 1.2 m \\
    24 & 2.4 m \\
    23 & 4.8 m \\
    \bottomrule
\end{tabular}

%% file: tables/param_theta_time.tex
\begin{tabular}{cl}
    \toprule
    parameter $\theta_{time}$ & time distance $\Theta_{time}$ \\
    \midrule
    32 & 1 s \\
    26 & 1 min \\
    24 & 4 min \\
    22 & 17 min \\
    \bottomrule
\end{tabular}

%% file: tables/psi-ex-balanced.tex
\begin{tabular}{c c c} 
\hline \hline
size $N$ (bytes) & \begin{tabular}{c}
{\bf Intel SGX} \\ execution (ms) / \\communication (MB)
\end{tabular} & \begin{tabular}{c}
{\bf OT} \cite{rindal2017malicious} \\ execution (ms) / \\communication (MB)
\end{tabular} \\
\hline
$10^3$ (16 KB) & 38 / 0.016 & 35 / 2 \\
$10^4$ (0.16 MB) & 52 / 0.16& 207 / 22 \\
$10^5$ (1.6 MB) & 153 / 1.6  & 2389 / 235 \\
$10^6$ (16 MB) & 1552 / 16 & 27110 / 2482 \\
$5.0\times10^6$ (80 MB) & 121526 / 80 & 154826 / 12502 \\
\hline \hline
\end{tabular}
\vspace{3pt}
\caption{PSI comparison: cryptography-based vs secure-hardware-based; execution time (balanced)}

%% file: tables/psi-ex-unbalanced.tex
\begin{tabular}{c c c c} 
\hline \hline
size $N$ (bytes) & size $n$ & 
\begin{tabular}{c}
{\bf Intel SGX} \\ execution (ms) / \\communication (MB)
\end{tabular} & 
{\bf HE \cite{chen2017fast} \footnotemark[5]} \\
\hline
$2^{16}$ (1 MB) & 5535 & 77 / 0.089 & 600 / 2.6  \\
$2^{16}$ (1 MB) & 11041 & 73 / 0.17 & 1300 / 4.1  \\
$2^{20}$ (17 MB) & 5535 & 72 / 0.089 & 2200 / 5.6  \\
$2^{20}$ (17 MB) & 11041 & 85 / 0.17 & 4000 / 12.0  \\
$2^{24}$ (268 MB) & 5535 & 249 / 0.089 & 10600 / 11.0  \\
$2^{24}$ (268 MB) & 11041 & 424 / 0.17 & 16200 / 21.1  \\
\hline \hline
\end{tabular}
\vspace{3pt}
\caption{PSI comparison: cryptography-based vs secure-hardware-based; execution time (unbalanced)}

%% file: tables/accuracy_dataset.tex
\begin{tabular}{|l|l|l|}
\hline
               & client query data & server infection data \\ \hline
\textbf{NY}    & $20160 \times 100$       & $20160 \times 1000$          \\ \hline
\textbf{Kinki} & $1440 \times 100$        & $1440 \times 14000$         \\ \hline
\textbf{Tokyo} & $1440 \times 100$         & $1440 \times 14000$         \\ \hline
\end{tabular}
\caption{Data set size used in accuracy evaluation. For example, New York's client query data has 100 clients and each of them has 20160 trajectory point records (e.g., every minute for 2 weeks), totally 2016000 records for client query data.}

%% file: tables/accuracy_normal.tex
\begin{tabular}{|l|l|l|l|l|l|l|l|l|}
\hline
                                              & $\theta_{geo}$ & $\theta_{time}$ & TP     & TN      & FP & FN     \\ \hline
\multirow{3}{*}{\textbf{NY}} 
                                              & 21         & 21               & 49\% (985635) & 37\% (745317)  & 0\% (326)  & 14\% (284722) \\
                                              & 24         & 22          & 29\% (575880) & 61\% (1224748) & 0\% (196)  & 11\% (215176) \\
                                              & 25         & 25          & 9\% (178467) & 80\% (1611751) & 0\% (246)  & 11\% (225536) \\ \hline
\textbf{Kinki}                    & 24         & 22          & 1\% (1007)   & 99\% (141409)  & 0\% (0)  & 0\% (45)     \\ \hline
\textbf{Tokyo}                    & 24         & 22          & 78\% (112459) & 20\% (28145)   & 0\% (2)  & 1\% (1885) \\ \hline
\end{tabular}
\caption{{\itshape st-PSI} can achieve high accuracy for all trajectory dataset, but cause false negatives.}

%% file: tables/accuracy_accurate.tex
\begin{tabular}{|l|l|l|l|l|l|l|}
\hline
                                              & $\theta_{geo}$ & $\theta_{time}$ & TP      & TN      & FP     & FN \\ \hline
\multirow{3}{*}{\textbf{NY}} & 21         & 21          & 63\% (1270303) & 26\% (530334)  & 11\% (215309) & 0\% (54) \\ 
                                              & 24         & 22          & 39\% (791054)  & 55\% (1104383) & 6\% (120561) & 0\% (2) \\ 
                                              & 25         & 25          & 20\% (404003)  & 74\% (1489927) & 6\% (122070) & 0\% (0)  \\ \hline
\textbf{Kinki}                    & 24         & 22          & 1\% (1052)    & 99\% (141380)  & 0\% (29)     & 0\% (0)  \\ \hline
\textbf{Tokyo}                    & 24         & 22          & 80\% (114313)  & 19\% (26772)   & 1\% (1375)   & 0\% (1)  \\ \hline
\end{tabular}
\caption{{\itshape nfp-stPSI} improves accuracy and remove false negatives, but cause false positive.}

%% file: tables/accuracy_obliv.tex
\begin{tabular}{|l|l|l|l|l|l|l|}
\hline
                                              & $\theta_{geo}$ & $\theta_{time}$ & TP      & TN      & FP     & FN \\ \hline
\textbf{NY}                    & 24         & 22          & 22\% (2206) & 70\% (6986) & 0\% (2)  & 8\% (806) \\ \hline
\end{tabular}
\caption{\cite{Reichert2020PrivacyPreservingCT} causes similar errors for subsampled NewYork dataset.}